\documentclass[11pt]{article}
\usepackage{graphicx, amssymb, amsmath, amsthm, bm, fullpage, authblk, dsfont}
\usepackage[colorlinks]{hyperref}
\usepackage[braket]{qcircuit}

\title{Efficiently constructing a quantum uniform superposition over bit strings near a binary linear code}

\author[1,2]{Edward Farhi\footnote{edwardfarhi@google.com}}
\author[1]{Stephen P. Jordan\footnote{stephenjordan@google.com}}
\affil[1]{\small{\it{Google Quantum AI}}}
\affil[2]{\small{\it{Massachusetts Institute of Technology}}}

\date{}

\newcommand{\id}{\mathds{1}}                      
\newcommand{\eq}[1]{(\ref{#1})}                   
\newcommand{\sect}[1]{\S\ref{#1}}                 
\renewcommand{\th}{^{\textrm{th}}}                

\newtheorem*{conv}{Convolution Theorem}
\newtheorem*{had}{Hadamard Transform of a Single Ball}
\newtheorem*{hadcode}{Hadamard of a Uniform Superposition of Codewords}
\newtheorem*{bdd}{Bounded Distance Decoding}
\newtheorem*{dual}{Dual Code}

\begin{document}

\bibliographystyle{unsrt}

\maketitle

\begin{abstract}
    We demonstrate that a high fidelity approximation to $\ket{\Psi_b}$, the quantum superposition over all bit strings within Hamming distance $b$ of the codewords of a dimension-$k$ linear code over $\mathbb{Z}_2^n$, can be efficiently constructed by a quantum circuit for large values of $n$, $b$ and $k$ which we characterize. We do numerical experiments at $n=1000$ which back up our claims. The achievable radius $b$  is much larger than the distance out to which  known classical algorithms can efficiently find the nearest codeword. Hence, these states cannot be prepared by quantum constuctions that require uncomputing to find the codeword nearest a string. Unlike the analogous states for lattices in $\mathbb{R}^n$, $\ket{\Psi_b}$ is not a useful resource for bounded distance decoding because the relevant overlap falls off too quickly with distance and known classical algorithms do better. Furthermore the overlap calculation can be dequantized. Perhaps these states could be used to solve other code problems. The technique used to construct these states is of interest and hopefully will have applications beyond codes.
\end{abstract}

\section{Introduction}
\label{sec:intro}

Given a probability distribution $p$ over the bit strings of length $n$ one can define a corresponding quantum state $\ket{p} = \sum_{\mathbf{x} \in \{0,1\}^n} \sqrt{p(\mathbf{x})} \ket{\mathbf{x}}$. The problem of preparing these states when $p$ can be efficiently sampled from classically was originally called qsampling in \cite{ATS03} but we call it Quantum Distributional State Construction, to emphasize that it is a problem of preparing coherent quantum states. Given $\ket{p}$ one can sample from the classical probability distribution $p$ by measuring $\ket{p}$ in the computational basis. QDSC can also be used to do things that classical sampling cannot. In particular, given two probability distributions $p$ and $q$ on $\{0,1\}^n$, there is no way in general to determine whether $p$ is close to $q$ (\emph{e.g.} in total variation distance) by drawing polynomially many samples from $p$ and $q$. In contrast, using polynomially many copies of $\ket{p}$ and $\ket{q}$ one can answer this question using the Hadamard test.

Here, we consider the problem of QDSC for the uniform distribution over all bit strings within radius $b$ of a binary linear code. Any binary linear code can be expressed in terms of a generator matrix $B \in \mathbb{Z}_2^{k \times n}$ as follows:
\begin{equation}
    \label{eq:generator}
    C = \{ \mathbf{x} B : \mathbf{x} \in \mathbb{Z}_2^k \}.
\end{equation}
The quantities $n$ and $k$ are known as the length and dimension of the code, respectively. There are $2^k$ codewords. (We are assuming that the rows of $B$ are linearly independent.) For a given linear code $C \subset \mathbb{Z}_2^n$ and a radius $b \in \{0,1,\ldots,n\}$  we will show how to efficiently produce the state
\begin{equation}
    \label{eq:psib}
    \ket{\Psi_b} = \mathcal{N} \sum_{\mathbf{c} \in C} \sum_{\mathbf{z} \in \mathbb{Z}_2^n}\Theta(b-|\mathbf{z}|) \ket{\mathbf{z} \oplus \mathbf{c}},
\end{equation}
where $| \mathbf{z} |$ denotes the Hamming weight of $\mathbf{z}$, and $\Theta$ is the unit step function:
\begin{equation}
    \label{eq:unitstep}
    \Theta(y)  = \left\{ \begin{array}{cl} 1 & \textrm{if } y \geq 0 \\ 0 & \textrm{if } y < 0 \end{array} \right. .
\end{equation}

For now, assume that $b$ is less than half the distance of the code. (The distance of the code is the minimum Hamming weight of the difference between any two codewords.) In this case, the state $\ket{\Psi_b}$ is a superposition of non-overlapping balls of radius $b$ centered at the codewords of $C$. The normalization factor is then
\begin{equation}
    \mathcal{N} = \frac{1}{\sqrt{2^k \mathrm{Vol}(b)}}
\end{equation}
where $\mathrm{Vol}(b)$ is the number of points contained within a ball of Hamming radius $b$:
\begin{equation}
    \mathrm{Vol}(b) = \sum_{j=0}^b \binom{n}{j}.
\end{equation}
If the balls overlap the normalization is more complicated.

The uniform probability distribution over all bit strings within Hamming distance $b$ of a binary linear code is easy to sample from classically. First, sample $\mathbf{x}$ uniformly from bit strings of length $k$ and sample $\mathbf{z}$ uniformly from bit strings of length $n$ and Hamming weight at most $b$. Then compute $\mathbf{x} B \oplus \mathbf{z}$. This suggests an approach to prepare the state $\ket{\Psi_b}$. First, prepare
\begin{equation}
    \Bigl( \frac{1}{\sqrt{2^k}} \sum_{\mathbf{x} \in \mathbb{Z}_2^k } \ket{\mathbf{x}} \Bigr) \Bigl( \frac{1}{\sqrt{\mathrm{Vol}(b)}} \sum_{\mathbf{z} \in \mathbb{Z}_2^n } \Theta(b-|\mathbf{z}|) \ket{\mathbf{z}} \Bigr) \ket{0}^{\otimes n}.
\end{equation}
Next, reversibly compute $\mathbf{x} B \oplus \mathbf{z}$ into the last register, yielding
\begin{equation}
    \frac{1}{\sqrt{2^k}} \sum_{\mathbf{x} \in \mathbb{Z}_2^k } \ket{\mathbf{x}} \ \frac{1}{\sqrt{\mathrm{Vol}(b)}} \sum_{\mathbf{z} \in \mathbb{Z}_2^n } \Theta(b-|\mathbf{z}|) \ket{\mathbf{z}} \ \ket{\mathbf{x} B \oplus \mathbf{z}}.
\end{equation}
Last, we need to uncompute $\mathbf{x}$ and $\mathbf{z}$ from the first two registers to obtain $\ket{\Psi_b}$. By assumption, $b$ is less than half the distance of the code. Therefore, the balls do not overlap and $\mathbf{x}$ and $\mathbf{z}$ are uniquely determined by $\mathbf{x} B \oplus \mathbf{z}$. However, determining $\mathbf{x}$ and $\mathbf{z}$ from $\mathbf{x} B \oplus \mathbf{z}$ is in general a computationally hard problem. Specifically, it is a bounded distance decoding problem for the code $C$ where the bound on distance is $b$.

\begin{bdd}
    We are given a linear code $C \subset \mathbb{Z}_2^n$ specified by its generator matrix, as in \eq{eq:generator}. We are given a bit string $\mathbf{v} \in \mathbb{Z}_2^n$ and are promised that $\mathbf{v}$ has Hamming distance at most $d$ from the nearest codeword: $\min_{\mathbf{c} \in C} | \mathbf{v} - \mathbf{c} | \leq d$. The bounded distance decoding problem is to find $\mathbf{c}$.
\end{bdd}

Bounded distance decoding for general binary linear codes can only be solved out to $d$ logarithmic in $n$ using known polynomial-time algorithms \cite{BK01, APY09}. Consequently the method just described is only capable of efficiently preparing the state $\ket{\Psi_b}$ for $b = O(\log n)$. Preparing $\ket{\Psi_b}$ out to larger radius by this direct approach fails due to an uncomputation barrier. In this work, we show that by taking a more intrinsically quantum approach we can surpass the uncomputation barrier and efficiently prepare $\ket{\Psi_b}$ out to $b$ much larger than the radius to which we can solve bounded distance decoding.

In \cite{ATS03,AR03} it was shown that the ability to prepare superpositions over points within Euclidean distance $b$ of the elements of a lattice in $\mathbb{R}^n$ would yield efficient quantum algorithms to solve classically-intractible instances of the Bounded Distance Decoding (BDD) problem over these lattices. No efficient quantum circuits for this state preparation task are known. The states $\ket{\Psi_b}$ that we prepare here are closely analogous to the states described in \cite{ATS03,AR03}, except over $\mathbb{Z}_2$ instead of $\mathbb{R}$ and using Hamming distance instead of Euclidean distance.

\begin{figure}
    \[
        \Qcircuit @C=1em @R=.7em {
        \lstick{(\ket{0}+\ket{1})/\sqrt{2}} & \qw & \ctrl{1} & \gate{H} & \meter \\
        \lstick{\ket{\Psi_b}} & {/} \qw  & \gate{T_{\mathbf{v}}} & \qw & \qw & }
    \]
    \caption{\label{fig:hadamard} This circuit implements the Hadamard test. The unitary $T_\mathbf{v}$ implements the translation by $\mathbf{v}$. That is, $T_{\mathbf{v}} \ket{\mathbf{x}} = \ket{\mathbf{x} \oplus \mathbf{v}}$. The probability that the final measurement yields outcome zero is $ \frac{1}{2} + \frac{1}{2} \bra{\Psi_b} T_{\mathbf{v}} \ket{\Psi_b}$.}
\end{figure}
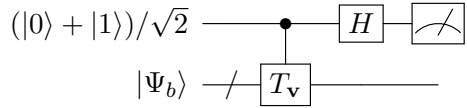

The state $\ket{\Psi_b}$ can be used as a resource for BDD on binary linear codes using the same approach that was proposed in \cite{ATS03, AR03} for lattices over $\mathbb{R}$. For $\mathbf{v} \in \mathbb{Z}_2^n$ let $T_\mathbf{v}$ be the corresponding unitary translation operator defined by $T_{\mathbf{v}} \ket{\mathbf{x}} = \ket{\mathbf{x} \oplus \mathbf{v}}$. If the overlap between $\ket{\Psi_b}$ and its translation by $T_{\mathbf{v}}$ is bigger than $0$ then $\mathbf{v}$ is within distance $2b$ of the lattice. Given $\ket{\Psi_b}$, one can use the Hadamard test to estimate the overlap $\bra{\Psi_b} T_{\mathbf{v}} \ket{\Psi_b}$, as illustrated in figure \ref{fig:hadamard}. This overlap is determined by the distance from $\mathbf{v}$ to the nearest codeword and decreases monotonically with this distance. (See \sect{sec:overlap}.) By iteratively flipping a bit of $\mathbf{v}$ and accepting the move if the distance is decreased one can find the nearest codeword.

The number of Hadamard tests required to detect that $\bra{\Psi_b} T_{\mathbf{v}} \ket{\Psi_b}$ is nonzero and hence that $\mathbf{v}$ is less than distance $2b$ from the nearest codeword is on the order of $1/|\bra{\Psi_b} T_{\mathbf{v}} \ket{\Psi_b}|^2$. Using amplitude amplification, this cost can be reduced quadratically~\cite{BHT02}. However, the overlap $|\bra{\Psi_b} T_{\mathbf{v}} \ket{\Psi_b}|$ falls rapidly with $|\mathbf{v}|$. Consequently, as we show in \sect{sec:bddimp}, using the states $\ket{\Psi_b}$ in this way beats brute force search but fails to outperform the classical algorithm Information Set Decoding. Furthermore, in \sect{sec:dequant}, we show that this approach can be dequantized.

It is interesting to see another example where the uncomputation barrier thwarts the naive approach to QDSC. As discussed in \cite{ATS03}, the graph isomorphism problem can be reduced to QDSC as follows. Given a graph $G$ with $n$ vertices, one wishes to prepare the uniform superposition over all of its permutations,
\begin{equation}
    \ket{\Phi_G} = \frac{1}{\sqrt{N(G)}} \sum_{\pi \in S_n} \ket{\pi(G)}
\end{equation}
where $N(G)$ is the appropriate normalization factor.
If $G$ and $G'$ are isomorphic graphs then $\langle \Phi_G | \Phi_{G'} \rangle = 1$. Otherwise $\langle \Phi_G | \Phi_{G'} \rangle = 0$. These cases are easily distinguished using the Hadamard test. Furthermore, the uniform probability distribution over permutations of a graph is easy to sample from classically by sampling a permutation at random and then applying it to $G$.

Now, consider the approach to quantum state preparation that is analogous to the above (failed) attempt to make $\ket{\Psi_b}$. For simplicity, we restrict attention to the case that $G$ has no automorphisms, so the number of distinct permutations is $n!$. The hoped for state construction process would be:
\begin{eqnarray}
    \frac{1}{\sqrt{n!}} \sum_{\pi \in S_n} \ket{\pi} \ket{G} & \to & \frac{1}{\sqrt{n!}} \sum_{\pi \in S_n} \ket{\pi} \ket{ \pi(G)} \\
    & \to & \frac{1}{\sqrt{n!}} \sum_{\pi \in S_n} \ket{\mathrm{identity}} \ket{\pi(G)}.
\end{eqnarray}
The second step of unentangling the first register from the second by reverting it to the identity is achievable if, with knowledge of $G$, $\pi$ can be computed from $\pi(G)$. In this case, one  computes $\pi$ from $\pi(G)$ reversibly in superposition and applies $\pi^{-1}$ to the first register. However, computing $\pi$ from $\pi(G)$ is exactly the graph isomorphism problem. So we have not made any progress.

In \cite{ATS03} reductions are also given from quadratic residuosity and discrete logarithm to QDSC. In the case of discrete logarithm, the uncomputation barrier arises just as in the above two examples. In the case of quadratic residuosity, the uncomputation task that arises is to uncompute a superposition rather than an individual bit string. These examples show that quantum approaches to several hard problems are thwarted by uncomputation barriers. Our results in this paper provide an example of how an uncomputation barrier can be surmounted.

\section{Preliminaries}

We now give the background necessary for understanding our construction of $\ket{\Psi_b}$. Our procedure begins by constructing a state over the dual code so let us review what the dual is.

\begin{dual} Given a code $C \subset \mathbb{Z}_2^n$, the dual code $C^\perp$ is defined as
\begin{equation}
    C^{\perp} = \{ \mathbf{d} \in \mathbb{Z}_2^n : \mathbf{d} \cdot \mathbf{c} = 0 \quad \forall \mathbf{c} \in C\},
\end{equation}
where $\mathbf{d} \cdot \mathbf{c} = \bigoplus_{i=1}^n d_i c_i$ is the $\mathbb{Z}_2^n$ inner product.
\end{dual}
\noindent
The dual to a code of length $n$ and dimension $k$ has length $n$ and dimension $n-k$. We call the $n-k$ by $n$ generator matrix of the dual code $B_{\perp}$. Given $B$ we can efficiently construct $B_{\perp}$, as discussed in appendix \ref{app:systematic}. 

On the quantum side each codeword and each dual codeword can be viewed as a computational basis state in a $2^n$-dimensional space. The normalized uniform superposition of the $2^k$ codewords is related to the normalized uniform superposition of the $2^{(n-k)}$ dual codewords by the Hadamard transform.  Given any function $f:\mathbb{Z}_2^n \to \mathbb{C}$, its Hadamard transform $\tilde{f}:\mathbb{Z}_2^n \to \mathbb{C}$ is defined by
\begin{equation}
    \tilde{f}(\mathbf{x}) = \frac{1}{\sqrt{2^n}} \sum_{\mathbf{y} \in \mathbb{Z}_2^n} (-1)^{\mathbf{x} \cdot \mathbf{y}} f(\mathbf{y}).
\end{equation}
This transform can be applied to the amplitudes of a quantum state by applying the Hadamard gate $H = \frac{1}{\sqrt{2}} \left[ \begin{array}{cc} 1 & 1 \\ 1 & -1 \end{array} \right]$ to each qubit:

\begin{equation}
    H^{\otimes n} \sum_{\mathbf{x} \in \mathbb{Z}_2^n} f(\mathbf{x}) \ket{\mathbf{x}} = \sum_{\mathbf{x} \in \mathbb{Z}_2^n} \tilde{f}(\mathbf{x}) \ket{\mathbf{x}}.
\end{equation}

\begin{hadcode}
Given a uniform superposition over a code $C$, its Hadamard transform is the uniform superposition over the dual code $C^\perp$. That is:
\begin{equation}
    \label{eq:cperp}
    \mathrm{If } \quad f(\mathbf{x}) = \frac{1}{\sqrt{|C|}} \sum_{\mathbf{c} \in C} \delta_{\mathbf{x},\mathbf{c}} \quad \mathrm{ then } \quad 
    \tilde{f}(\mathbf{x}) = \frac{1}{\sqrt{|C^\perp|}} \sum_{\mathbf{d} \in C^\perp} \delta_{\mathbf{x},\mathbf{d}}.
\end{equation}
\end{hadcode}

\begin{proof}
Writing out the Hadamard transform yields
\begin{equation}
    \tilde{f}(\mathbf{x}) =\frac{1}{\sqrt{2^n}} \frac{1}{\sqrt{|C|}} \sum_{\mathbf{c} \in C}  (-1)^{\mathbf{x} \cdot \mathbf{c}}. 
\end{equation}
For $\mathbf{x} \in C^\perp$
\begin{equation}
    \frac{1}{\sqrt{|C|}} \sum_{\mathbf{c} \in C} (-1)^{\mathbf{x} \cdot \mathbf{c}} = \sqrt{|C|}.
\end{equation}
Recall that  $|C| = 2^k$ and accordingly $|C^\perp| = 2^{n-k}$. Thus $\tilde{f}(\mathbf{x}) = \frac{1}{\sqrt{|C^\perp|}}$ for any $\mathbf{x} \in C^\perp$. For all $\mathbf{x} \notin C^{\perp}$ one must have $\tilde{f}(\mathbf{x}) = 0$ since any additional nonzero amplitudes would yield norm greater than one, which cannot occur because the Hadamard transform is unitary.
\end{proof}

Our algorithm for preparing $\ket{\Psi_b}$ begins by constructing a preliminary state $\ket{\widetilde{\Psi}_b}$ and then applying the quantum Hadamard transform $H^{\otimes n}$. The state $\ket{\widetilde{\Psi}_b}$ is the Hadamard transform of the target state  $\ket{\Psi_b}$, 
\begin{equation}
    \label{eq:psihad}
    \ket{\widetilde{\Psi}_b} = H^{\otimes n} \ket{\Psi_b},
\end{equation}
so we see that acting on this state with the Hadamard transformation gives the target state.

By a slightly involved calculation which we defer to \sect{sec:hadamard},
\begin{equation}
    \label{eq:psiperp}
    \ket{\widetilde{\Psi}_b} = N \sum_{\mathbf{d} \in C^\perp} K_b^{n-1}(|\mathbf{d}| -1 ) \ \ket{\mathbf{d}},
\end{equation}
where $K_j^n(x)$ is the Krawtchouk polynomial defined by
\begin{equation}
    \label{eq:Krawtchouk}
    K_j^n(x) = \sum_{r=0}^j \binom{x}{r} \binom{n-x}{j-r} (-1)^r,
\end{equation}
and $N$ is the normalization factor. If the balls do not overlap then
\begin{equation}
    \label{eq:Ndef}
    N = \frac{1}{\sqrt{2^{n-k} \mathrm{Vol}(b)}},
\end{equation}
which is also shown in section  \sect{sec:hadamard}.

If we can construct $\ket{\widetilde{\Psi}_b}$ we can act on it with the Hadamard transform and get our desired state,

\begin{equation}
    \ket{\Psi_b} = H^{\otimes n} \ket{\widetilde{\Psi}_b}.
\end{equation}
So our task now is to give an efficient construction of  $\ket{\widetilde{\Psi}_b}$.
Suppose we can make the state
\begin{equation}
    \label{eq:unsigned}
    \ket{|\widetilde{\Psi}_b|} = N \sum_{\mathbf{d} \in C^\perp} \big| K_b^{n-1}(|\mathbf{d}| -1) \big| \ \ket{\mathbf{d}}.
\end{equation}
For a given dual codeword $\mathbf{d} \in C^\perp$ it is easy to compute the sign of the amplitude $K_b^{n-1}(|\mathbf{d}| -1)$ and apply it using phase kickback, as described in appendix \ref{app:kickback}, to make the state \eq{eq:psiperp} from \eq{eq:unsigned}. So the task now is to make this unsigned state. 

The unsigned state can be described in terms of a probability distribution as follows. Let $B_{\perp} \in \mathbb{Z}^{k^\perp \times n}$ be the generator matrix for $C^\perp$. Again $k^\perp = n - k$. Then,
\begin{equation}
\label{eq:target}
    \ket{|\widetilde{\Psi}_b|} = \sum_{\mathbf{u} \in \mathbb{Z}_2^{k^\perp}} \sqrt{p(\mathbf{u})} \ket{\mathbf{u}B_{\perp} },
\end{equation}
where $p(\mathbf{u})$ is the normalized probability distribution over $\mathbb{Z}_2^{k^\perp}$ given by
\begin{equation}
    \label{eq:pb}
    p(\mathbf{u}) = N^2  \left[ K_b^{n-1}(|\mathbf{u} B_{\perp}| -1) \right]^2
\end{equation}
States of the form $\sum_{\mathbf{x}} \sqrt{p(\mathbf{x})} \ket{\mathbf{x}}$ can in some cases be prepared efficiently using the method of conditional rotations, described in the next section.

\section{Conditional Rotations by Metropolis Monte Carlo}
\label{sec:rotations}

Given a probability distribution $p(u_1, \ldots, u_{k^\perp})$, the method of conditional rotations, introduced in \cite{Z98} and further developed in \cite{KM01, GR02}, can be used to  efficiently prepare the state
\begin{equation}
    \sum_{\mathbf{u} \in \mathbb{Z}_2^{k^\perp}} \sqrt{p(\mathbf{u})} \ket{\mathbf{u}}
\end{equation}
provided certain marginal conditional probabilities can be efficiently computed. Here, our goal is to prepare this state in the case that $p(\mathbf{u})$ is as given in \eq{eq:pb}. Once we have done this, it is straightforward to perform the transformation $\sum_{\mathbf{u}} \sqrt{p(\mathbf{u})} \ket{\mathbf{u}} \to \sum_{\mathbf{u}} \sqrt{p(\mathbf{u})} \ket{\mathbf{u} B_\perp}$, using classical reversible computation, to get the desired state \eq{eq:target}. This is spelled out in \sect{sec:alg}. Our method for computing the necessary conditional marginal probabilities is Metropolis Monte Carlo.

The method of conditional rotations works as follows. Start with $k^\perp$ qubits initialized to the zero state, $\ket{0}^{\otimes k^\perp}$. Suppose there is an efficient way to calculate the marginals $p(u_1=0)$ and $p(u_1=1)$ which are the probabilities that the first bit is $0$ or $1$. Then perform a rotation on the first qubit, obtaining
\begin{equation}
\left(\sqrt{p(u_1=0)} \ket{0} + \sqrt{p(u_1=1)} \ket{1} \right) \ket{0}^{\otimes (k^\perp-1)}.
\end{equation}
Assuming there is an efficient way to compute subsequent conditional marginals, perform a rotation on the second qubit conditioned on the value of the first qubit, obtaining 
\begin{equation}
\begin{array}{l} \left[ \sqrt{p(u_1=0)} \ket{0} \left(\sqrt{p(u_2=0|u_1=0)} \ket{0}+\sqrt{p(u_2=1|u_1=0)} \ket{1}\right) + \right. \vspace{10pt} \\
  \left. \sqrt{p(u_1=1)} \ket{1} \left(\sqrt{p(u_2=0|u_1=1)} \ket{0} + \sqrt{p(u_2=1|u_1=1)} \ket{1}\right) \right] \ket{0}^{\otimes(k^\perp-2)} \end{array}.
\end{equation}
Similarly, rotate the third qubit conditioned on the first two according to $p(u_3|u_2,u_1)$ and so on until all the bits are rotated. At the end of this process
one is left with the state 
\begin{equation}
\sum_{u_1 \ldots u_{k^\perp}} a(u_1, \ldots, u_{k^\perp}) \ket{u_1 \ldots u_{k^\perp}}
\end{equation}
where
\begin{eqnarray}
\label{exact}
a(u_1,\ldots,u_n) & = & \sqrt{p(u_1)} \sqrt{p(u_2|u_1)} \times \ldots \times \sqrt{p(u_{k^\perp}|u_{k^\perp-1},\ldots,u_1)} \label{eq:markov1} \\
 & = & \sqrt{p(u_1,\ldots,u_{k^\perp})} \label{eq:markov2}
\end{eqnarray}
as desired.

For this process to be efficient, it must be possible to efficiently evaluate all the necessary conditional marginal probabilities in which the first $m$ bits are fixed and the last $k^\perp-m$ bits are summed over, keeping track of how often bit $m+1$ is $0$ or $1$ to get the marginal. Classically precomputing these sums would be infeasible because there are exponentially many of them. Instead, these evaluations are  done in superposition by reversible circuits. Thus, the key question is whether these conditional marginal probabilities can each be computed by an efficient classical algorithm.

The classical algorithm that we propose for estimating these conditional probabilities is the method of Markov chain Monte Carlo, using the Metropolis rule. We find that the necessary Markov chains converge remarkably rapidly, despite the exponentially large state space of codewords in $C^\perp$ for certain values of $b$ and $k$.  We discuss the conditions that guarantee good convergence in \sect{sec:largeb}.

Consider first the Markov chain used to estimate the rotation probability for the first bit. At each step in the Markov chain our random walker is located at some string $\mathbf{u} \in \mathbb{Z}_2^{k^\perp}$. We then choose one bit of $\mathbf{u}$ unformly at random and flip it, yielding $\mathbf{u}'$. Since we are going for \eq{eq:pb} we then evaluate:
\begin{equation}
    f(\mathbf{u}') = \left[ K^{n-1}_b(|\mathbf{u'} B_{\perp}|-1)\right]^2.
\end{equation}
If  $f(\mathbf{u}') \geq f(\mathbf{u})$ this move is accepted and the walker moves from $\mathbf{u}$ to $\mathbf{u}'$. Otherwise, the move is accepted with probability $f(\mathbf{u}')/f(\mathbf{u})$. This is the Metropolis rule. Under mild conditions, this is guaranteed to converge to the limiting normalized distribution proportional to $f$.

One obtains a sequence of bit strings $\mathbf{u}, \mathbf{u}', \ldots \in \mathbb{Z}_2^{k_\perp}$ by running this Markov chain. The fraction of bit strings sampled in which the first bit is $1$ is used as the marginal distribution for the first qubit rotation. 
For the $m\th$ qubit rotation the same procedure is followed except that the first $m$ bits of $\mathbf{u}$ are frozen to specific values, and only flips of remaining $k^\perp-m$ bits are proposed in the random walk, thereby sampling from the necessary conditional distribution. This is done reversibly in superposition over values that the first $m$ bits are fixed to. Reversible implementation of the Markov chain is to be carried out using a reversible pseudorandom number generator.

\section{Quantum State Preparation Algorithm for \ket{\Psi_b}}
\label{sec:alg}

We now present pseudocode for the construction of $\ket{\Psi_b}$, which is an example of the Quantum Distributional State Construction problem.  It relies on sampling by Metropolis Monte Carlo which we described in the last section.
The key equation \eq{eq:psiperp} is established later so if you take it on faith hopefully you have enough to follow the pseudocode.

Given a generator matrix for $C$, we construct $\ket{\Psi_b}$ using the following steps.

\begin{enumerate}

    \item Precompute $K_b^{n-1}(h-1)$ for $h=0, \ldots, n$ so they need not be evaluated with the quantum device.  Note that they do not depend on the code.

    \item Starting from the generator matrix for $C$, compute $B_{\perp}$, the $k^\perp \times n$ generator matrix for $C^\perp$. This can be done in polynomial time, see for example appendix \ref{app:systematic}.

    \item Initialize a register of $k^\perp$ qubits and second register of $n$ qubits to the all zero state,
    \[
        \ket{0}^{\otimes k^\perp} \ket{0}^{\otimes n}.
    \]

    \item Use the method of conditional rotations, estimated using the Metropolis Monte Carlo method, to construct a weighted superposition over the coefficient vectors $\mathbf{u}$,
    \[
        \to N \sum_{\mathbf{u} \in \mathbb{Z}_2^{k^\perp}} \big | K_b^{n-1} ( |\mathbf{u} B_{\perp}|-1) \big | \ \ket{\mathbf{u}} \ket{0}^{\otimes n}.
    \]
    where $N$ is the normalization constant.
    Note that our method automatically produces a normalized state so there is no need to know $N$ in advance.

    \item Reversibly compute the corresponding codewords of $C^\perp$ into the second register,
    \[
        \to N \sum_{\mathbf{u} \in \mathbb{Z}_2^{k^\perp}} \big | K_b^{n-1}(|\mathbf{u}B_{\perp}|-1) \big | \ \ket{\mathbf{u}} \ket{\mathbf{u} B_{\perp}}.
    \]

    \item Uncompute the first register, returning it to $\ket{0}^{\otimes k^\perp}$. We do this after having classically precomputed the right-inverse $B_{\perp}^{-1}$ for the (non-square) matrix $B_{\perp}$, that is, $B_{\perp} B_{\perp}^{-1} = \id$.  Then in the first register add $(\mathbf{u} B_{\perp}) B_{\perp}^{-1}$ to $\mathbf{u}$. This yields
    \[
    \to N \sum_{\mathbf{u} \in \mathbb{Z}_2^{k^\perp}} \big | K_b^{n-1}(|\mathbf{u}B_{\perp}|-1) \big | \ \ket{\mathbf{0}} \ket{\mathbf{u} B_{\perp}}.
    \]
    This can be rewritten as
    \[
        = N \sum_{\mathbf{d} \in C^\perp} \big | K_b^{n-1}(|\mathbf{d}|-1) \big | \ \ket{\mathbf{d}},
    \]
    where we have discarded the first register for clarity.

    \item Use phase kickback (\emph{c.f.} appendix \ref{app:kickback}) to put in the correct sign of $K_b^{n-1}(|\mathbf{d}|-1)$,
    \[
        \to N \sum_{\mathbf{d} \in C^\perp} K_b^{n-1}(|\mathbf{d}|-1) \ \ket{\mathbf{d}}.
    \]

    \item Perform the quantum Hadamard transform $H^{\otimes n}$. As we show next in \sect{sec:hadamard}, this yields the desired final state,
    \[
        \to \mathcal{N} \sum_{\mathbf{c} \in C} \sum_{\mathbf{z} \in \mathbb{Z}_2^n} \Theta(b-|\mathbf{z}|) \ \ket{\mathbf{z} \oplus \mathbf{c}} = \ket{\Psi_b}.
    \] 
\end{enumerate}

The fidelity with which the final state is produced depends entirely on the precision to which the Metropolis Monte Carlo method can estimate the conditional probabilities defining the qubit rotations. This is step 4. All other steps of the algorithm are exact.

\section{Deriving the form of $\ket{\widetilde{\Psi}_b}$}
\label{sec:hadamard}

In this section we collect and prove some facts regarding Hadamard transforms and then use them to derive \eq{eq:psiperp}.

\begin{conv}
    \label{prop:convolution}
    \begin{equation}
        \label{eq:conv}
        H^{\otimes n} \frac{1}{\sqrt{2^n}}\sum_{\mathbf{z} \in \mathbb{Z}_2^n}  \sum_{\mathbf{y} \in \mathbb{Z}_2^n} f(\mathbf{y}) g(\mathbf{z} \oplus \mathbf{y}) \ket{\mathbf{z}} = \sum_{\mathbf{x} \in \mathbb{Z}_2^n} \tilde{f}(\mathbf{x}) \tilde{g}(\mathbf{x}) \ket{\mathbf{x}}
    \end{equation}
    where $\tilde{f}$ is the Hadamard transform of $f$
    \begin{equation}
        \tilde{f}(\mathbf{x}) = \frac{1}{\sqrt{2^n}} \sum_{\mathbf{y} \in \mathbb{Z}_2^n} (-1)^{\mathbf{x} \cdot \mathbf{y}} f(\mathbf{y})
    \end{equation}
    and $\tilde{g}$ is the Hadamard transform of $g$.
\end{conv}

\begin{proof}
\begin{equation}
    H^{\otimes n} \frac{1}{\sqrt{2^n}} \sum_{\mathbf{z} \in \mathbb{Z}_2^n}  \sum_{\mathbf{y} \in \mathbb{Z}_2^n} f(\mathbf{y}) g(\mathbf{z} \oplus \mathbf{y}) \ket{\mathbf{z}} = \frac{1}{2^n} \sum_{\mathbf{x},\mathbf{y},\mathbf{z} \in \mathbb{Z}_2^n} (-1)^{\mathbf{x} \cdot \mathbf{z}} f(\mathbf{y}) g(\mathbf{z} \oplus \mathbf{y} ) \ket{\mathbf{x}}
\end{equation}
Let $\mathbf{w} = \mathbf{z} \oplus \mathbf{y}$. Then the previous expression can be rewritten as

\begin{align}
    &= \frac{1}{2^n} \sum_{\mathbf{x},\mathbf{y},\mathbf{w} \in \mathbb{Z}_2^n} (-1)^{\mathbf{x} \cdot (\mathbf{w} \oplus \mathbf{y})} f(\mathbf{y}) g(\mathbf{w}) \ket{\mathbf{x}} \\
    &= \sum_{\mathbf{x} \in \mathbb{Z}_2^n} \Bigl( \frac{1}{\sqrt{2^n}} \sum_{\mathbf{y} \in \mathbb{Z}_2^n} (-1)^{\mathbf{x} \cdot \mathbf{y}} f(\mathbf{y}) \Bigr) \Bigl( \frac{1}{\sqrt{2^n}} \sum_{\mathbf{w} \in \mathbb{Z}_2^n} (-1)^{\mathbf{x} \cdot \mathbf{w}} g(\mathbf{w}) \Bigr) \ket{\mathbf{x}} \\
    &= \nonumber \sum_{\mathbf{x} \in \mathbb{Z}_2^n} \tilde{f}(\mathbf{x}) \tilde{g}(\mathbf{x}) \ket{\mathbf{x}}. \qedhere 
\end{align}
\end{proof}

\begin{had}
    \label{prop:ball}
    Given a radius $b \in \{0,1,2,\ldots,n\}$, let $g:\mathbb{Z}_2^n \to \mathbb{C}$ be
    \begin{equation}
        g(\mathbf{x}) = \Theta(b-|\mathbf{x}|),
    \end{equation}
    where $\Theta$ is the unit step function as defined in \eq{eq:unitstep}. Then, the Hadamard transform of $g$ is
    \begin{equation}
        \label{eq:gtilde}
        \tilde{g}(\mathbf{x}) = \frac{1}{\sqrt{2^n}} K_b^{n-1}(|\mathbf{x}|-1),
    \end{equation}
    where $K_b^{n-1}$ is the Krawtchouk polynomial as defined in \eq{eq:Krawtchouk}.
\end{had}

\begin{proof}
To compute the Hadamard transform of a ball, first rewrite
$g(\mathbf{y}) = \Theta(b-|\mathbf{y}|)$ as $g(\mathbf{y}) = \sum_{j = 0}^b \delta_{j,|\mathbf{y}|}$. Then

\begin{equation}
\tilde{g}(\mathbf{x}) = \frac{1}{\sqrt{2^n}} \sum_{j=0}^b \sum_{\mathbf{y} \in \mathbb{Z}_2^n} (-1)^{\mathbf{x} \cdot \mathbf{y}} \delta_{j,|\mathbf{y}|}.
\end{equation}
The summand $(-1)^{\mathbf{x} \cdot \mathbf{y}} \delta_{j,|\mathbf{y}|}$ depends on $\mathbf{x}$ and $\mathbf{y}$ only through $r$, the number of bits which are equal to 1 in both $\mathbf{x}$ and $\mathbf{y}$, and $s$, the number of bits which are equal to 1 in $\mathbf{y}$ but not $\mathbf{x}$. Thus, 
\begin{equation}
\tilde{g}(\mathbf{x}) = \frac{1}{\sqrt{2^n}} \sum_{j=0}^b \sum_{r = 0}^{|\mathbf{x}|} \sum_{s = 0}^{n-|\mathbf{x}|} \binom{|\mathbf{x}|}{r} \binom{n - |\mathbf{x}|}{s} (-1)^r \delta_{j,r+s}
\end{equation}
\begin{equation}
= \frac{1}{\sqrt{2^n}} \sum_{j=0}^b \left( \sum_{r=0}^j \binom{|\mathbf{x}|}{r} \binom{n - |\mathbf{x}|}{j-r} (-1)^r \right).
\end{equation}
The inner expression is recognizable as a Krawtchouk polynomial, which is defined by 
\begin{equation}
K_j^n(x) = \sum_{r=0}^j \binom{x}{r} \binom{n-x}{j-r} (-1)^r
\end{equation}
thus we can write 
\begin{equation}
\tilde{g}(\mathbf{x}) = \frac{1}{\sqrt{2^n}} \sum_{j=0}^b K_j^n(|\mathbf{x}|).
\end{equation}
Using the identity 
\begin{equation}
\label{eq:ident}
\sum_{j=0}^b K_j^n(x) = K_b^{n-1}(x-1),
\end{equation}
which we derive in appendix \ref{app:ident}, we can further simplify our expression to 
\[
\tilde{g}(\mathbf{x}) = \frac{1}{\sqrt{2^n}} K_b^{n-1}(|\mathbf{x}|-1). \qedhere
\]
\end{proof}

With these facts in hand, we can derive \eq{eq:psiperp} as follows. Examination of \eq{eq:psib} shows that $\ket{\Psi_b}$ is proportional to a convolution in the form of \eq{eq:conv}. Namely,
\begin{equation}
    \label{eq:psib2}
    \ket{\Psi_b} = \sqrt{\frac{2^n}{\mathrm{Vol}(b)}} \frac{1}{\sqrt{2^n}} \sum_{\mathbf{z} \in \mathbb{Z}_2^n}  \sum_{\mathbf{y} \in \mathbb{Z}_2^n} f(\mathbf{y}) g(\mathbf{z} \oplus \mathbf{y}) \ket{\mathbf{z}},
\end{equation}
where
\begin{equation}
    f(\mathbf{x}) = \frac{1}{\sqrt{2^k}} \sum_{\mathbf{c} \in C} \delta_{\mathbf{c},\mathbf{x}}
\end{equation}
and
\begin{equation}
    g(\mathbf{x}) = \Theta(b-|\mathbf{x}|).
\end{equation}
The convolution theorem therefore tells us that
\begin{equation}
    \label{eq:penultimate}
    H^{\otimes n} \ket{\Psi_b} = \sqrt{\frac{2^n}{\mathrm{Vol}(b)}} \sum_{\mathbf{x} \in \mathbb{Z}_2^n} \tilde{f}(\mathbf{x}) \tilde{g}(\mathbf{x}) \ket{\mathbf{x}}.
\end{equation}
Now the Hadamard transform of $f(\mathbf{x})$ is given by \eq{eq:cperp} and  the Hadamard transform of $g(\mathbf{x})$ is given by  \eq{eq:gtilde} so plugging in gives

\begin{equation}
    \label{eq:psiperpp}
    \ket{\widetilde{\Psi}_b}  = H^{\otimes n} \ket{\Psi_b}= N \sum_{\mathbf{d} \in C^\perp} K_b^{n-1}(|\mathbf{d}| -1 ) \ket{\mathbf{d}}
\end{equation}
where $N$ is given as
\begin{equation}
    N = \frac{1}{\sqrt{2^{n-k} \mathrm{Vol}(b)}}.
\end{equation}
So we have established \eq{eq:psiperp}.

\section{Impact of Imperfect Convergence on Fidelity}
\label{sec:imperfect}

 In \sect{sec:convergence} we show computer experiments to see how well a Metropolis Monte Carlo algorithm does at sampling from a Krawtchouk distribution at $n=1000$ with $10^{10}$ samples.  In this section we quantify the effect of sampling with a finite number of samples versus the ideal limiting distribution. We calculate the fidelity between the ideal quantum state and one that arises from imperfect sampling assuming that the only source of error is the imperfect sampling. We do assume that the sampled distribution is uniform over codewords of a given Hamming weight, as it is in the ideal case. The result is \eq{eq:impover} in terms of the ideal and sampled distributions and is evaluated in specific cases in the next section.

Let
\begin{equation}
    \label{eq:hdef}
    \ket{h} = \frac{1}{\sqrt{W(h)}} \sum_{\substack{\mathbf{d} \in C^\perp \\ |\mathbf{d}| = h}} \ket{\mathbf{d}},
\end{equation}
where $W(h)$ is the number of codewords in $C^\perp$ with Hamming weight $h$. Note that we are not assuming anything about $W(h)$. Then
\begin{equation}
    \label{eq:orthonormal}
    \langle h | h' \rangle = \delta_{h,h'}.
\end{equation}
By \eq{eq:unsigned},
\begin{equation}
    \label{eq:psiperph}
    \ket{|\widetilde{\Psi}_b|} = N \sum_{h=0}^n \sqrt{W(h)} \big| K^{n-1}_b(h-1) \big| \ \ket{h}.
\end{equation}
The probability distribution over Hamming weights defined by this state is
\begin{equation} \label{eq:pideal}
  p_{\textrm{ideal}}(h) = \frac{\left[ K^{n-1}_b(h-1) \right]^2 W(h)}{\sum_{h=0}^n \left[ K^{n-1}_b(h-1) \right]^2 W(h)}.
\end{equation}
We can correspondingly rewrite \eq{eq:psiperph} as
\begin{equation}
    \ket{|\widetilde{\Psi}_b|} = \sum_{h=0}^n \sqrt{p_{\textrm{ideal}}(h)} \ket{h}.
\end{equation}
Let $p_{\mathrm{sam}}(h)$ be the probability distribution over Hamming weights achieved from the product of conditional probabilities estimated by Markov chains, as in \eq{eq:markov1} and \eq{eq:markov2}. Assuming no other sources of error, the unsigned state resulting from conditional rotations would be
\begin{equation}
    \label{eq:psiperpsample}
    \ket{ | \widetilde{\Psi}_b^{\mathrm{sam}} |} = \sum_{h=0}^n \sqrt{p_{\mathrm{sam}}(h)} \ket{h}.
\end{equation}
Here is where we assumed that the distribution over sampled dual codewords at fixed Hamming weight is uniform.
By \eq{eq:orthonormal}, the inner product between 
$\ket{ |\widetilde{\Psi}_b^{\mathrm{sam}}|}$ and $\ket{ |\widetilde{\Psi}_b | }$ is
\begin{equation}
    \left \langle | \widetilde{\Psi}_b^{\mathrm{sam}} | \right| \left. |\widetilde{\Psi}_b| \right \rangle = \sum_{h=0}^n \sqrt{p_{\mathrm{sam}}(h)} \sqrt{p_{\mathrm{ideal}}(h)}.
\end{equation}
The minus signs imposed by phase kickback are always exact, and so the signed state will have this same inner product
\begin{equation}
    \left\langle \widetilde{\Psi}_b^{\mathrm{sam}} \right| \left. \widetilde{\Psi}_b \right\rangle = \sum_{h=0}^n \sqrt{p_{\mathrm{sam}}(h)} \sqrt{p_{\mathrm{ideal}}(h)}.
\end{equation}
Let $\ket{\Psi_b^{\mathrm{sam}}}$ be the approximation to $\ket{\Psi_b}$ obtained by taking the Hadamard transform of $\ket{\widetilde{\Psi}_b^{\mathrm{sam}}}$,
\begin{equation}
    \label{eq:psia}
    \ket{\Psi_b^{\mathrm{sam}}} = H^{\otimes n} \ket{\widetilde{\Psi}_b^{\mathrm{sam}}}.
\end{equation}
Now $| \Psi_b \rangle $ is the Hadamard of $\ket{\widetilde{\Psi}_b}$ so the fidelity of the ideal state  with the sampled state is
\begin{equation}
    \label{eq:impover}
    \langle \Psi_b^{\mathrm{sam}} | \Psi_b \rangle = \sum_{h=0}^n \sqrt{p_{\mathrm{sam}}(h)} \sqrt{p_{\mathrm{ideal}}(h)}.
\end{equation}
By doing numerical experiments where we classically sample to get $p_{\mathrm{sam}}(h)$ we can estimate the quantum fidelity between the less than perfect quantum state and the ideal.

\section{Empirical Convergence of Metropolis Monte Carlo}
\label{sec:convergence}

To see if the Monte Carlo algorithm converges with not too many samples we consider the Markov chain used in the first step of the method of conditional rotations. That is, none of the bits of $\mathbf{u}$ are fixed and the state space consists of all $2^{k^\perp}$ bit strings $\mathbf{u}$. Any subsequent Monte Carlo step with $m$ bits of $\mathbf{u}$ frozen is essentially a version of the same problem with $m$ rows of the generator matrix for $C^\perp$ removed. As we will see in \sect{sec:largeb}, failures of convergence can arise when one or both of the ball radius $b$ and the Hamming distance $f$ of the step size in the random walk are too large. The step size $f$ is the Hamming weight of a row of the generator matrix $B_\perp$. As $B_\perp$ is in systematic form, these rows have Hamming weight approximately $k/2$. In the $m\th$ random walk, we have removed $m$ of these rows. This leaves $b$ unchanged and $f$ still has an expected size of $k/2$.  So we expect that if the first step converges, the subsequent ones do as well.

For our examples we  use random codes. Because this state space is exponentially large, it is not practical to verify the convergence of the Metropolis Markov chain by directly comparing the sampling frequencies to \eq{eq:pb}. Instead, we measure the frequency at which elements of $C^\perp$ with given Hamming weight are sampled. If the algorithm is converged, these sampling frequencies should be proportional to
\begin{equation}
    W(h) \left[ K_b^{n-1}(h-1)\right]^2
\end{equation}
where $W(h)$ is the number of codewords in $C^\perp$ with Hamming weight $h$. It is well known that random linear codes have a weight distribution which is close to binomial
\begin{equation}
    W(h) \simeq \frac{2^{k^\perp}}{2^n} \binom{n}{h}
\end{equation}
so we take
\begin{equation}
 \label{eq:ph}
p_{\mathrm{ideal}}(h) \propto \binom{n}{h} \left[ K_b^{n-1}(h-1)\right]^2.
\end{equation}
We do not need the explicit form of the normalization for what follows.

\begin{figure}
\begin{center}
\includegraphics[width=0.85\textwidth]{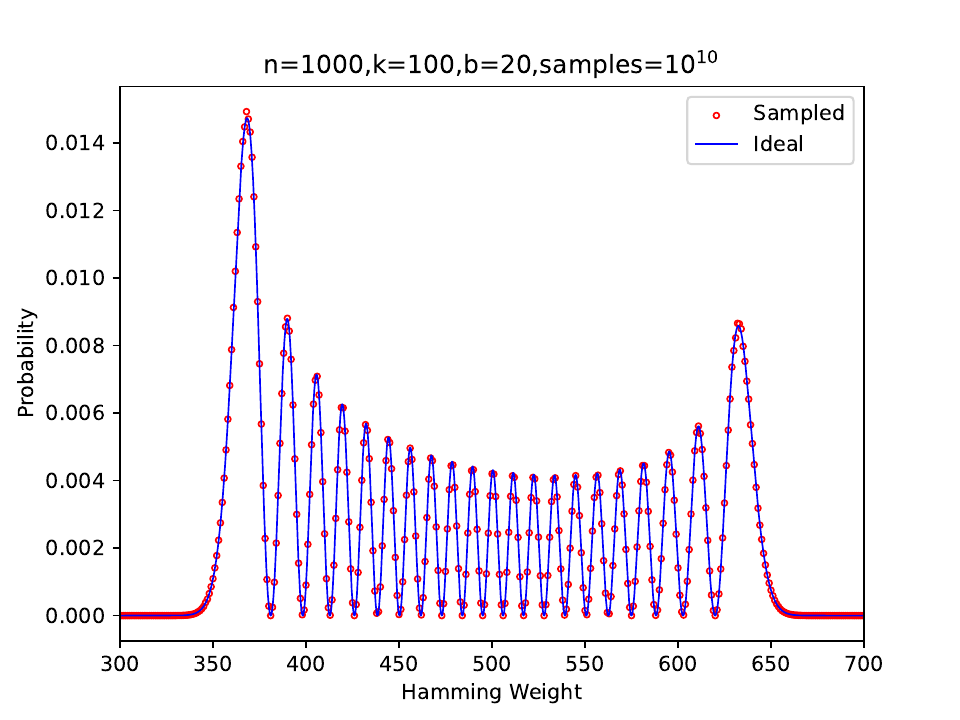}
\caption{\label{fig:good_full} Comparison between  $p_{\mathrm{ideal}}(h)$ and the probability distribution in Hamming weight produced by Metropolis Monte Carlo sampling on a random linear code with $n=1000$ and $k^\perp = n-k = 900$ basis vectors. The ball radius is $b=20$. The Metropolis random walk was run for $10^{10}$ steps. The fidelity between the ideal quantum state and the sampled quantum state is $0.999995$. We plot only from Hamming weight 300 to 700 because outside of that range the sampled probabilities are zero.}
\end{center}
\end{figure}

We want to see if sampling can produce the distribution in \eq{eq:ph}.  We show two numerical examples. The first is at $n=1000$, $k=100$, $b=20$ and is shown in figure \ref{fig:good_full}. 
The results show close agreement between the measured sampling frequencies as a function of Hamming weight and the prediction of \eq{eq:ph}. This close agreement suggests both that the Metropolis algorithm is well converged and that the binomial distribution is a good approximation to $W(h)$.

Figure \ref{fig:good_full} shows that  $p_{\mathrm{ideal}}(h)$ is very small outside of a limited range of Hamming weights. In the example shown $p_{\mathrm{ideal}}(h)$ is very small for Hamming weights less than roughly 340 or greater than roughly 660. This is significant, because finding codewords of low (or high) Hamming weight is an approximate version of the lightest codeword problem which is classically hard. Algorithms  that require  sampling from dual codewords of very low Hamming weight are unlikely to be successful.  In this example, for our task of constructing the state \eq{eq:psib}, we succeed without going near low Hamming weight dual codewords.  

 For a linear code, we can estimate the minimum Hamming weight of any nonzero codeword. As shown in \cite{BF02}, if a generator matrix is chosen uniformly at random then with high probability the resulting code will have minimum distance $D$ saturating the Gilbert-Varshamov bound. That is, $D$ is obtained by solving
\begin{equation}
\label{eq:gvdist}
2^k = \frac{2^n}{\textrm{Vol}(D)} = \frac{2^n}{\sum_{j=0}^D \binom{n}{j}}.
\end{equation}
In our example the dual is a random code, and we estimate that minimum Hamming weight of any nonzero dual codeword is approximately 14. We see that we are not sampling near the smallest dual codewords and we do not need to.

\begin{figure}
\begin{center}
\includegraphics[width=0.85\textwidth]{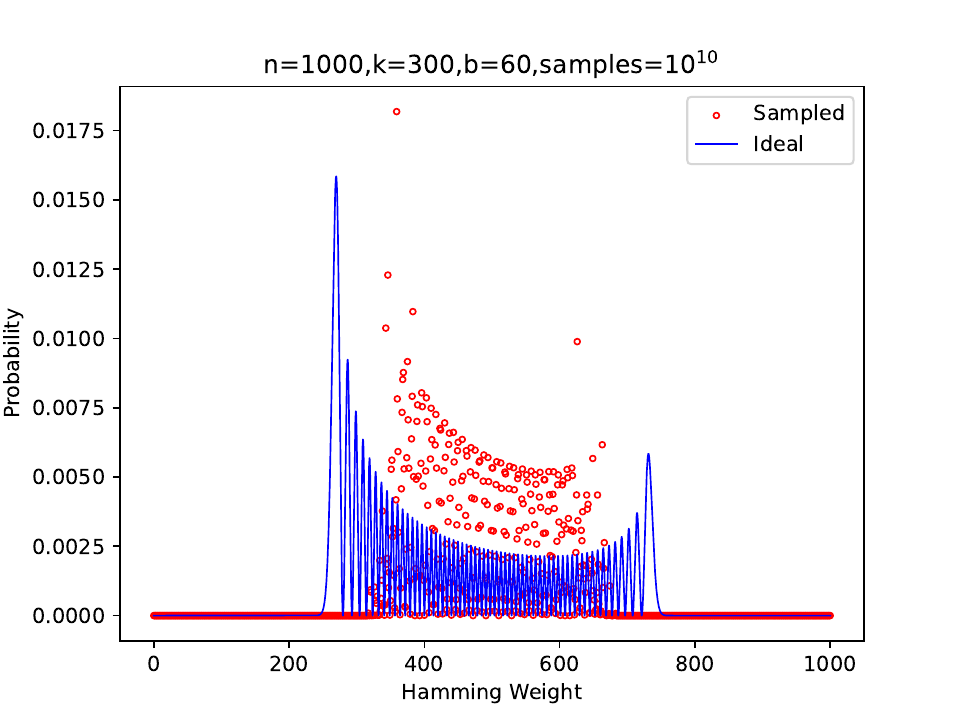}
\caption{\label{fig:bad_full} Comparison between  $p_{\mathrm{ideal}}(h)$ and the probability distribution in Hamming weight produced by Metropolis Monte Carlo sampling on a random linear code with $n=1000$ and $k^\perp = n-k = 700$ basis vectors. The ball radius is $b = 60$.  The Metropolis random walk was run for $10^{10}$ steps. Incomplete convergence is observed because the region of nonnegligible support for $p_{\mathrm{ideal}}(h)$ extends to dual codewords of  small and  large Hamming weight which are inaccessible to the random walk. Still the fidelity between the ideal quantum state and sampled quantum state is $0.67$ because the sampling in the middle range is very accurate as is shown in figure \ref{fig:bad_middle}.}
\end{center}
\end{figure}

\begin{figure}
\begin{center}
\includegraphics[width=0.85\textwidth]{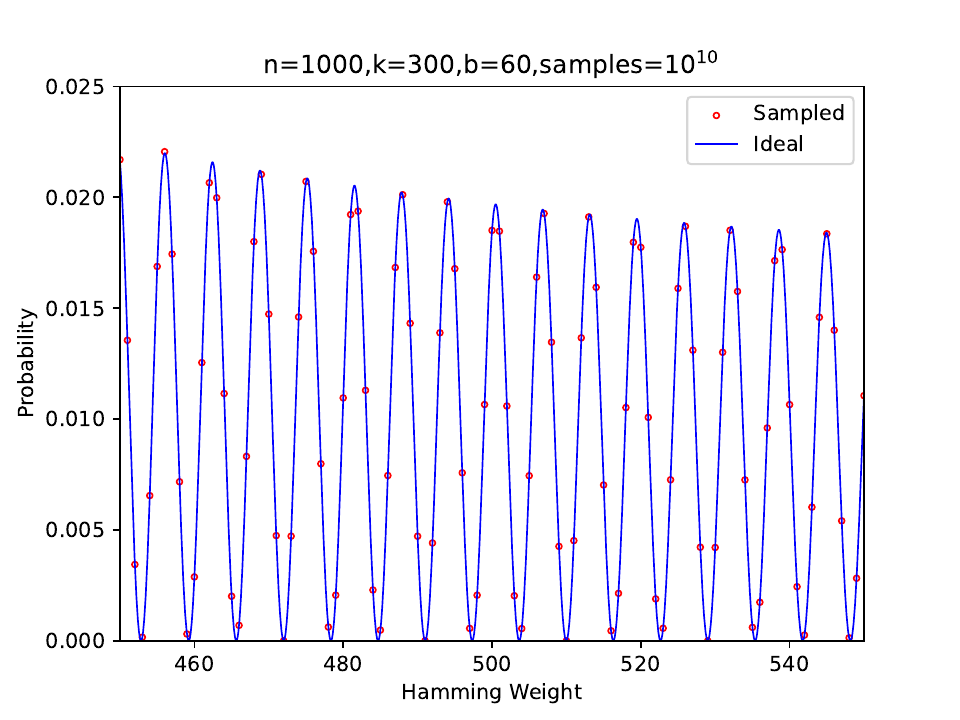}
\caption{\label{fig:bad_middle} Here is a blowup of the central region of figure \ref{fig:bad_full}. In figure \ref{fig:bad_full} the sampled frequencies for these Hamming weights are higher than in $p_{\mathrm{ideal}}(h)$  because the sampled frequencies on the low and high Hamming weight tails are too small so the central region gets amplified to maintain the normalization. Here we show only the range $450 \leq h \leq 550$ and renormalize both the ideal and sampled distributions so that these probabilities in this region both add to one.}
\end{center}
\end{figure}

By choosing larger $k$ and $b$ we can construct examples in which the support of $p_{\mathrm{ideal}}(h)$ extends to dual codewords of  low and high Hamming weight which take exponential time to reach. In such cases, the Markov chain converges nicely to the correct relative probabilities for the dual codewords of Hamming weight in the middle range but the outer tails are cut off. We illustrate this in figure \ref{fig:bad_full} where $n=1000$, $k=300$, and $b=60$.  In figure \ref{fig:bad_middle} we show the central region but we rescale the two distributions so they both sum to one in this limited range.  Note the excellent agreement between the sampled and the ideal in the middle range.  This results in high fidelity between the ideal quantum state and the sampled quantum state. A quantitative analysis of which values of $n,k,b$ are expected to yield fully-converged Markov chains as in figure \ref{fig:good_full} versus which are expected to yield cut-off distributions as in figure \ref{fig:bad_full} is given in \sect{sec:largeb}.

Note that, even in this second example ($n=1000$, $k=300$, and $b=60$), the ideal  distribution over dual codewords has negligible probability at Hamming weight below approximately 263, whereas the shortest nonzero dual codeword for a random code of these parameters is, with high probability, approximately 65. So achieving full convergence would not require sampling  near the shortest nonzero codeword.

Another phenomenon that we frequently observe in the unconverged regime is that some codewords just at the boundaries of the regions of accessible Hamming weights are sampled with frequencies that appear to be anomalously high.  An example of this phenomenon can be seen in figure \ref{fig:bad_full} in which there are a few outlier red points showing sampling frequencies substantially higher than expected.

By the results of \sect{sec:imperfect}, namely \eq{eq:impover}, we can compute the fidelity between a state arising from the sampled probability distributions and the ideal state, under the assumption that the sampling of codewords of a given Hamming weight is uniform as in the ideal case. We are ignoring the error that occurs from successive conditional rotations.  For the well converged example at $n=1000,k=100,b=20$ from figure \ref{fig:good_full} we obtain a fidelity of 0.999995. For the example at $n=1000,k=300,b=60$ shown in figure \ref{fig:bad_full} we obtain fidelity 0.67. These overlaps are high and show that in practice we should be able to make the desired states with good fidelity.

\clearpage

\section{Analysis of Convergence of Metropolis Monte Carlo}
\label{sec:largeb}

In this section we derive a criterion to predict which values of $n,k,b$ will yield good convergence, and which will not. We then compare these predictions against numerical experiments.

Based on the form of $p_{\mathrm{ideal}}(h)$ illustrated in figure \ref{fig:good_full}, we identify two circumstances that cause concern regarding the convergence of our Markov chain to its limiting distribution. The first we will show is actually not problematic. The second can prevent full convergence for some values of $n,k,b$ that we fully characterize.

First, one expects the random walk to fail if $p_{\mathrm{ideal}}(h)$ contains regions of high probability separated by a region of very low probability with width greater than the Hamming distance of the trial steps. We would expect failure in this case because the Metropolis walk would need to go from one region to the other only by stepping across a region of low probability. The Metropolis walk will not accept moves into very low probability regions so they will create ``bottlenecks'' for the Metropolis walk.

As shown in figure \ref{fig:good_full},  $p_{\mathrm{ideal}}(h)$ oscillates, and has many points of low probability. If the Metropolis walk were only allowed to make moves of Hamming distance one, these low probability codewords would create bottlenecks, which might cause the Metropolis walk to fail. However, the walk actually moves by taking steps of Hamming distance one on the coefficient vector $\mathbf{u}$. A step $\mathbf{u} \to \mathbf{u}'$ where $|\mathbf{u}' - \mathbf{u}| = 1$ corresponds to a step from the dual codeword $\mathbf{u}B_\perp$ to the dual codeword $\mathbf{u}'B_\perp$. That is, the new dual codeword is obtained by adding one row of $B_\perp$ to the current dual codeword. The distance of the move will be the Hamming weight of the row. When $B_\perp$ is in systematic form, as defined in appendix \ref{app:systematic}, the typical Hamming weight of a row is is approximately $(1/2)(n-k^\perp) = k/2$. By taking steps of Hamming distance $k/2$, the Metropolis walk can easily step over regions of very low probability as is confirmed in our numerical examples.

It is known that the Krawtchouk polynomial $K_b^n(x)$ has $b$ real zeros which all lie in the range $n/2-\sqrt{b (n-b)} \leq x \leq n/2+\sqrt{b (n-b)}$ \cite{IS98,QW04}. So the average spacing between the  zeros is $2\sqrt{(n-b)/b}$ and numerically we see that the zeros are roughly uniformly spaced. Many of the values of $x$ where the exact zeros occur are not at integers but as we can see from figure \ref{fig:bad_middle} at the integer values near the exact zeros the polynomial is tiny.
As discussed above, a bottleneck could occur if there were a ``trough'' in which $p_{\mathrm{ideal}}(h)$ is consistently low over some region whose width is larger than the steps taken by the random walker. So  we can see that unless $2\sqrt{(n-b)/b} \gg k/2$, the oscillations in $p_{\mathrm{ideal}}(h)$ are of short enough wavelength that any troughs will be much narrower than the step length for the walker and this type of bottleneck will not occur.  Accordingly in  our numerical experiments we do not see this failure mode.

The second circumstance in which we might expect the Markov chain to fail to converge is when the distribution $p_{\mathrm{ideal}}(h)$ assigns nonnegligible probability to codewords of Hamming weight near either zero or $n$, which become exponentially hard to reach by the sampling process. (Note that this is not the problem of finding short codewords which are difficult to discover.  It is a problem of failure to sample.)
If a random walker is sitting at a bit string of very low Hamming weight, a move that is proposed in the Metropolis algorithm which flips roughly $k/2$ bits will be much more likely to raise the Hamming weight than lower it. And the reverse holds at very high Hamming weight. If the edges of the nonnegligible region for $p_{\mathrm{ideal}}(h)$ extend beyond these entropic barriers, then we expect the random walk to converge incompletely. After any reasonable number of moves in the random walk, the number of times the walker reaches the regions beyond the entropic barrier will remain tiny, thereby cutting off the tails of the desired distribution $p_{\mathrm{ideal}}(h)$.  We see this in figure  \ref{fig:bad_full}.

Unlike bottlenecks, which we believe never occur, this second failure mode can occur when $b$ or $k$ is large. We can quantify this as follows. Consider a random walker at bit string $\mathbf{d}$ that has Hamming weight $h$. At a given step of the Metropolis algorithm a move is proposed by adding a random row of $B_\perp$, which will have Hamming weight $f \simeq k/2$ since $B_\perp$ is in systematic form. We model this as flipping $f$ bits of $\mathbf{d}$ chosen uniformly at random.  Let $j$ be the number of ones in $\mathbf{d}$ that flip and accordingly the number of zeros that flip is $f-j$. The Hamming weight of the flipped string is $(h-j)+(f-j)$ so to decrease the Hamming weight we need $j > f/2$. Counting the number of ways that this can happen we get that the probability for the proposed move to decrease the Hamming weight is
\begin{equation}
    \label{eq:barrier}
    P_{\mathrm{down}}(h) = \frac{\sum_{j=\lceil f/2 \rceil}^f \binom{h}{j} \binom{n-h}{f-j}}{\binom{n}{f}}.
\end{equation} 
If $h$ becomes too small relative to $n$ and $f$ then this probability becomes exponentially small. The value of $h$ below which $P_{\mathrm{down}}$ becomes insurmountably small relative to the number of steps in the Markov chain defines an entropic barrier which the random walker is unlikely to cross. One also has the analogous situation at $h$ close to $n$ for which  $P_{\mathrm{up}}$ becomes insurmountably small.

The existence of this entropic barrier only prevents convergence if the region to which $p_{\mathrm{ideal}}(h)$ assigns nonnegligible probability extends outside this entropic barrier. Using the results of \cite{IS98} on the asymptotics of Krawtchouk polynomials, we find that the region on which $p_{\mathrm{ideal}}(h)$ is nonnegligible is
\begin{equation}
    \label{eq:kraw_barrier}
    n/2-\sqrt{b(n-b)} \leq h \leq n/2+\sqrt{b(n-b)}.
\end{equation}
Since this region is symmetric about $n/2$, the region of nonnegligible $p_{\mathrm{ideal}}(h)$ extends above the high-Hamming-weight entropic barrier  for the same values of $n,k,b$ in which it extends below the low-Hamming-weight barrier. Hence, given $n,k,b$ we can predict whether the Markov chain will converge by solving $P_{\mathrm{down}}(h) = \epsilon$ for $h$ using some appropriately small $\epsilon$. If the resulting value $h_{\mathrm{barrier}}(\epsilon)$ satisfies $h_{\mathrm{barrier}}(\epsilon) < n/2-\sqrt{b(n-b)}$ then the Markov chain can be expected to converge, and otherwise not.

\begin{figure}
\begin{center}
\includegraphics[width=0.85\textwidth]{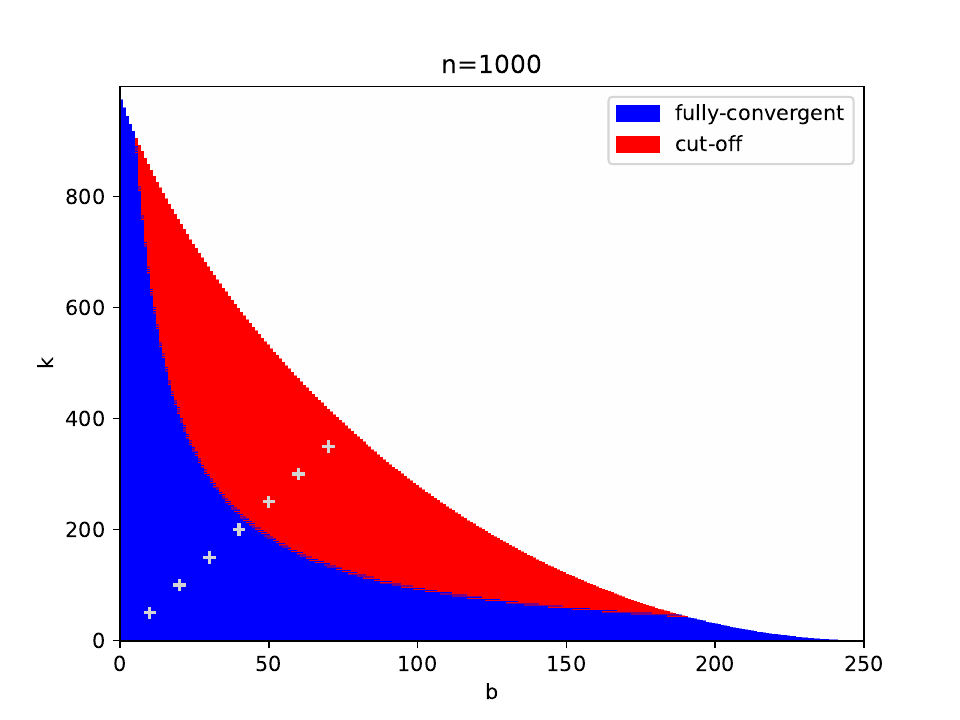}
\caption{\label{fig:region_diagram} At $n=1000$ we show the region (blue) in the $b-k$ plane in which we predict our Markov chain to efficiently converge and the region (red) in which we expect the sampled distribution to be resemble the ideal distribution with its low and high Hamming weight tails cut off. The boundary between the the red and blue regions is described after \eq{eq:kraw_barrier}. The white region corresponds to overlapping balls ($2b$ greater than the code distance), which we do not consider in this paper. The grey crosses illustrate the location of the Monte Carlo experiments plotted in figure \ref{fig:mcexp}.} 
\end{center}
\end{figure}

Let us now compare this prediction against computer experiments. The location of the entropic barrier is not highly sensitive to the choice of $\epsilon$ so we semi-arbitrarily set $\epsilon=10^{-6}$. Using this criterion, for any fixed $n$ we can map out the $k-b$ plane, classifying each point into one of three categories. The first category is where the left entropic barrier $h_{\mathrm{barrier}}(\epsilon)$ is below the Krawtchouk edge at $n/2-\sqrt{b(n-b)}$. In this case we expect efficient convergence across the full range of nonnegligible $p_{\mathrm{ideal}}(h)$. The second region is where the entropic barrier is above the Krawtchouk edge. In this case we expect the sampled distribution $p_{\mathrm{sam}}(h)$ to look like $p_{\mathrm{ideal}}(h)$ with the outer tails cut off. The third region is where the ball radius exceeds the minimum distance of the code $C$, that is $2b > D$. This is the regime of overlapping balls, which we do not consider in this paper. These regions are plotted for $n=1000$ in figure \ref{fig:region_diagram}.

We next examine a slice through these regions along the line $k=5 \times b$. We choose a set of seven evenly-spaced points along this line, and at each one generate ten random codes and carry out our Markov chain for each one. The analysis above predicts that the first four points, which lie in the fully-convergent region, should yield sampled distributions over Hamming weights with high fidelity to the ideal distribution. The last three points, which lie in the cut-off region, should have fidelity falling off due to an increasing fraction of the tails of the distribution being cut off. Two of these seven points, namely $k=100,b=20$ and $k=300,b=60$ are used in \sect{sec:convergence} to illustrate in more detail the behavior in the convergent and cut-off regions, respectively.

\begin{figure}
\begin{center}
\includegraphics[width=0.8\textwidth]{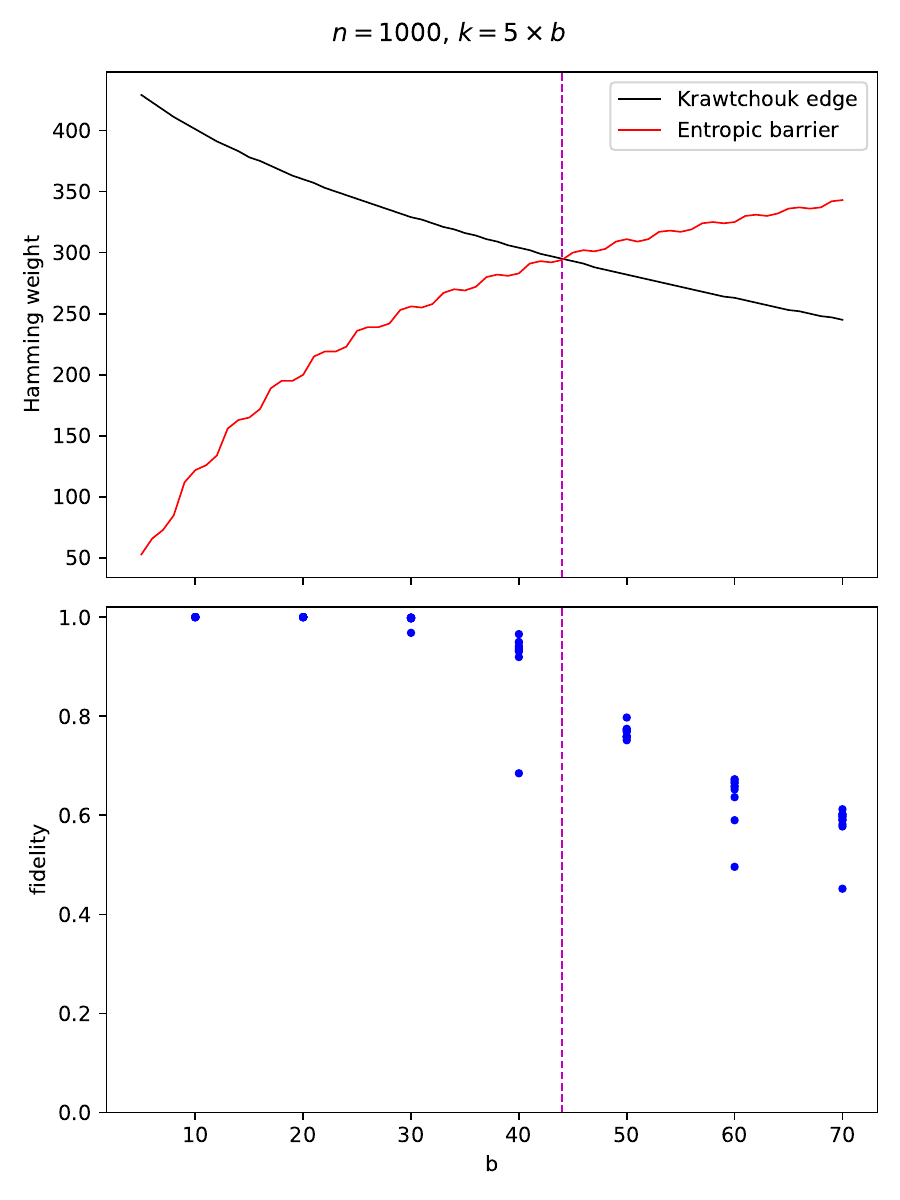}
\caption{\label{fig:mcexp} Here, at $n=1000$ we take a slice through the $k-b$ plane illustrated in figure \ref{fig:region_diagram} by setting $k=5 \times b$. For $b=10, 20, ... 70 $ we run ten trials of the Markov chain each with $10^{10}$ steps. Then we compute the fidelity $\sum_{h=0}^n \sqrt{p_{\mathrm{sam}}(h)} \sqrt{p_{\mathrm{ideal}}(h)}$. The first four values of $b$ are in the region where we expect full convergence, and we observe high fidelity. The last three are in the cut-off region, and we observe the fidelity drop off.}
\end{center}
\end{figure}

The results are shown in figure \ref{fig:mcexp}, which show good qualitative agreement with these predictions.
For our trials at the three points in the cut-off region, the fidelity, as calculated with \eq{eq:impover}, is greater than 0.5, as shown in figure \ref{fig:mcexp}. This is higher than one might naively expect since a look at figure  \ref{fig:bad_full}  suggests that the sampled and ideal distributions are not close.  However inside of the entropic barrier where $p_{\mathrm{sam}}$ is not tiny, the two distributions are in fact proportional, that is, one is scale factor times the other.  See figure \ref{fig:bad_middle} where both have been normalized over the plotted region.  The quantum state overlap  \eq{eq:impover} has support only inside the entropic region where the two distributions are proportional and the fidelity is then the square root of the scale factor which is not very small. 

\clearpage

\section{Overlap of $\ket{\Psi_b}$ with $T_{\mathbf{v}} \ket{\Psi_b}$}
\label{sec:overlap}

Our original motivation for finding a way to construct $\ket{\Psi_b}$ is that the overlap of this state with its translation by $\mathbf{v}$ gives information about how close $\mathbf{v}$ is to the  lattice.  We do not claim that we have an  efficient quantum algorithm for BDD. Still we want to understand how this overlap depends on how far $\mathbf{v}$ is from the lattice and in this section we derive an expression for the ideal state overlap $\bra{\Psi_b} T_{\mathbf{v}} \ket{\Psi_b}$. This determines the runtime needed to solve BDD using Hadamard tests, which we discuss in \sect{sec:bddimp}.

Let $\boldsymbol\delta$ be the displacement of $\mathbf{v}$ from the nearest codeword. That is
\begin{equation}
    \label{eq:bolddelta}
    \boldsymbol\delta = \mathbf{v} \oplus \mathbf{c}
\end{equation}
where $\mathbf{c}$ is the element of the code $C$ that minimizes $| \mathbf{v} \oplus \mathbf{c}|$.
If each ball in $\ket{\Psi_b}$ overlaps with only one ball from $T_{\mathbf{v}} \ket{\Psi_b}$ then $\bra{\Psi_b} T_{\mathbf{v}} \ket{\Psi_b}$ is equal to the overlap between a pair of single normalized  balls displaced by $\boldsymbol \delta$. This is because there are $2^k$ overlapping ball pairs and there is the same factor in the denominator from the overall normalization. We denote the overlap of a single ball with its translate by $\boldsymbol\delta$ as $\mathcal{A}(\boldsymbol \delta)$.

\begin{figure}
    \begin{center}
    \includegraphics[width=0.5\textwidth]{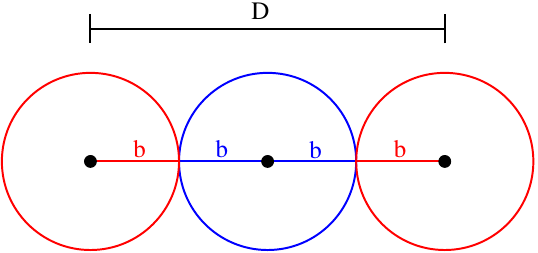}
    \caption{\label{fig:4r} The red circles represent balls from $T_{\mathbf{v}} \ket{\Psi_b}$. Their centers are separated by Hamming distance $D$, which is the minimum distance of the code $C$. The blue circle represents a ball from $\ket{\Psi_b}$. All balls have radius $b$. Here we illustrate that if $4b < D$ then it is impossible for any ball from $\ket{\Psi_b}$ to overlap more than one ball from $T_{\mathbf{v}} \ket{\Psi_b}$. The closest it can come to this is if $\delta = 2b$ which is shown. This is illustrated using Euclidean distance, but the same holds for Hamming distance, as it also obeys the triangle inequality.}
    \end{center}
\end{figure}

Since the triangle inequality holds for Hamming distance, one can observe that if $2b + |\boldsymbol \delta|$ is less than the minimum distance of the code $C$ then it is guaranteed that each ball in $\ket{\Psi_b}$ will overlap with at most one ball from $T_{\mathbf{v}} \ket{\Psi_b}$. Furthermore, if $4b$ is less than the minimum distance of $C$ then this will hold for all $\mathbf{v}$, as illustrated in figure \ref{fig:4r}. In these cases $\bra{\Psi_b} T_{\mathbf{v}} \ket{\Psi_b} = \mathcal{A}(\boldsymbol \delta)$.  
In our first example in \sect{sec:convergence} we have $n=1000, k=100$ and  $b=20$. We estimate the length of the smallest codeword is $320$ so we are safely in the regime of $4b < D$. For our next example at $n=1000, k=300$ and $b=60$ we estimate that the smallest codeword has length $192$ so for $\delta$ up to $72$ we have $2b + |\boldsymbol \delta| < D$ and so each ball in $\ket{\Psi_b}$ will overlap only one ball from $T_{\mathbf{v}} \ket{\Psi_b}$. 

By definition
\begin{equation}
    \mathcal{A}(\boldsymbol \delta) = \langle B_1 | B_2 \rangle
\end{equation}
where, for $j=1,2$
\begin{eqnarray}
    \ket{B_j} & = & \frac{1}{\sqrt{|B_j|}} \sum_{\mathbf{x} \in B_j} \ket{\mathbf{x}} \\
    B_j & = & \{ \mathbf{x} \in \mathbb{Z}_2^n : |\mathbf{x} \oplus \mathbf{z}_j| \leq b \} \\
\end{eqnarray}
and $\mathbf{z}_1,\mathbf{z}_2$ are any pair of bit strings satisfying
\begin{equation}
    \mathbf{z}_1 \oplus \mathbf{z}_2 = \boldsymbol\delta.
\end{equation}
Now
\begin{equation}
    \label{eq:bigpicture}
    \bra{\Psi_b} T_{\mathbf{v}} \ket{\Psi_b} = \langle B_1 | B_2 \rangle = \frac{|B_1 \cap B_2|}{\sqrt{|B_1| \cdot |B_2|} } = \frac{|B_1 \cap B_2|}{ |B_1| },
\end{equation}
where we used that $|B_1| = |B_2|$.
Next, observe that
\begin{equation}
    \label{eq:b1}
    |B_1| = \sum_{j=0}^b \binom{n}{j}.
\end{equation}
To compute $|B_1 \cup B_2|$ we can assume without loss of generality that
\begin{eqnarray*}
  \mathbf{z}_1 & = & \overbrace{00\ldots000\ldots0}^{n} \\
  \mathbf{z}_2 & = & \underbrace{11\ldots1}_{\delta}\underbrace{00\ldots0}_{n-\delta}
\end{eqnarray*}
where $\delta = |\boldsymbol\delta|$, which is equal to the distance from $\mathbf{v}$ to the nearest codeword.

Starting from $\mathbf{z}_2$ one generates all elements of $B_2$ by flipping $s_1$ bits from the first $\delta$ and $s_2$ bits from the last $n-\delta$, where $s_1 + s_2 \leq b$. The number of ways to do this is $\binom{\delta}{s_1} \binom{n-\delta}{s_2}$. The Hamming weight of the resulting string is $\delta-s_1+s_2$. Therefore,
\begin{equation}
    \label{eq:intersection}
    | B_1 \cap B_2 | = \sum_{(s_1,s_2) \in P} \binom{\delta}{s_1} \binom{n-\delta}{s_2}
\end{equation}
where
\begin{equation}
    \label{eq:P}
    P = \{ (s_1,s_2) : s_1 \geq 0, s_2 \geq 0, s_1+s_2 \leq b \textrm{, and } \delta-s_1+s_2 \leq b \}.
\end{equation}
Putting it all together, we have
\begin{equation}
    \mathcal{A}(\boldsymbol \delta) = \frac{\sum_{(s_1,s_2) \in P} \binom{\delta}{s_1} \binom{n-\delta}{s_2}}{\sum_{j=0}^b \binom{n}{j}}
\end{equation}
with $P$ as given in \eq{eq:P}. In the case that each ball overlaps with at most one other,
\begin{equation}
    \label{eq:ideal_overlap}
    \bra{\Psi_b} T_{\mathbf{v}} \ket{\Psi_b} = \frac{\sum_{(s_1,s_2) \in P} \binom{\delta}{s_1} \binom{n-\delta}{s_2}}{\sum_{j=0}^b \binom{n}{j}},
\end{equation}
where $\delta$ is the Hamming distance from $\mathbf{v}$ to the nearest codeword.

\section{Dequantizing the calculation of $\bra{\Psi_b} T_{\mathbf{v}} \ket{\Psi_b}$}
\label{sec:dequant}

Here we show that there is a classical sampling procedure that one can run to estimate $\bra{\Psi_b} T_{\mathbf{v}} \ket{\Psi_b}$ by sampling a $\pm 1$ variable and averaging the result. The number of samples needed to distinguish this quantity from zero is of order $1/|\bra{\Psi_b} T_{\mathbf{v}} \ket{\Psi_b}|^2$, which has the same scaling as the number of repetitions of the Hadamard test that are needed in the quantum setting. The quantity $\bra{\Psi_b} T_{\mathbf{v}} \ket{\Psi_b}$ is exponentially small in the distance of $\mathbf{v}$ from the lattice so both the quantum and dequantized algorithms require time exponential in this distance to distinguish $\bra{\Psi_b} T_{\mathbf{v}} \ket{\Psi_b}$ from zero. The dequantization shown here is modeled on the dequantization results regarding lattices on $\mathbb{R}^n$ shown in \cite{AR05}.

To dequantize the estimation of $\bra{\Psi_b} T_{\mathbf{v}} \ket{\Psi_b}$ we first write $T_{\mathbf{v}}$ in terms of Pauli $X$ operators which flip bits where $\mathbf{v}$ is $1$,
\begin{equation}
T_{\mathbf{v}} = \prod_{i \ \mathrm{ s.t. } \ v_i=1} X_i .
\end{equation}
Next, we recall that $\ket{\Psi_b} = H^{\otimes n} \ket{\widetilde{\Psi}_b}$. Since $H X_i H = Z_i$ we have
\begin{eqnarray}
    \bra{\Psi_b} T_{\mathbf{v}} \ket{\Psi_b} & = & \bra{\Psi_b} \prod_{i \ \mathrm{ s.t. } \ v_i=1} X_i \ket{\Psi_b} \\
    & = & \bra{\widetilde{\Psi}_b} H^{\otimes n} \prod_{i \ \mathrm{ s.t. } \ v_i=1} X_i H^{\otimes n} \ket{\widetilde{\Psi}_b} \\
    & = & \bra{\widetilde{\Psi}_b} \prod_{i \ \mathrm{ s.t. } \ v_i=1} Z_i \ket{\widetilde{\Psi}_b}.
\end{eqnarray}
Recalling the expression \eq{eq:psiperp} for $\ket{\widetilde{\Psi}_b}$ this yields
\begin{equation}
     \bra{\Psi_b} T_{\mathbf{v}} \ket{\Psi_b} = N^2 \sum_{\mathbf{d} \in C^\perp} \left[ K_b^{n-1}(|\mathbf{d}|-1) \right]^2 \bra{\mathbf{d}} \prod_{i \ \mathrm{ s.t. } \ v_i=1} Z_i \ket{\mathbf{d}}.
\end{equation}
The quantity $N^2 \left[ K_b^{n-1}(|\mathbf{d}|-1) \right]^2$ is the probability distribution over dual codewords that we sample from in our Markov chain. See \eq{eq:pb}. In a slight abuse of notation we define
\begin{equation}
    p_{\mathrm{ideal}}(\mathbf{d}) = N^2 \left[ K_b^{n-1}(|\mathbf{d}|-1) \right]^2,
\end{equation}
where $N$ is the normalization factor \eq{eq:Ndef}. The previously defined probability $p_{\mathrm{ideal}}(h)$ over Hamming weights is related to this by
\begin{equation}
    \label{eq:ideal_relation}
    p_{\mathrm{ideal}}(h) = \sum_{\substack{\mathbf{d} \in C^\perp \\ |\mathbf{d}|=h}} p_{\mathrm{ideal}}(\mathbf{d}).
\end{equation}
For any given codeword, the quantity $\bra{\mathbf{d}} \prod_{i \ \mathrm{ s.t. } \ v_i=1} Z_i \ket{\mathbf{d}} \in \{+1,-1\}$ is easy to compute. Namely
\begin{equation}
    \bra{\mathbf{d}} \prod_{i \ \mathrm{ s.t. } \ v_i=1} Z_i \ket{\mathbf{d}} = (-1)^{\mathbf{d} \cdot \mathbf{v}}.
\end{equation}
Consequently,
\begin{equation}
    \label{eq:dequant}
    \bra{\Psi_b} T_{\mathbf{v}} \ket{\Psi_b} = \sum_{\mathbf{d} \in C^\perp} p_{\mathrm{ideal}}(\mathbf{d}) (-1)^{\mathbf{d} \cdot \mathbf{v}}.
\end{equation}

The formula \eq{eq:dequant} yields the following dequantization of our procedure for estimating $\bra{\Psi_b} T_{\mathbf{v}} \ket{\Psi_b}$. Use Metropolis Monte Carlo to sample dual codewords according to $p_{\mathrm{ideal}}$. For each dual codeword sampled, compute $(-1)^{\mathbf{d} \cdot \mathbf{v}}$. After $S$ samples, the average of this list of $\pm 1$ values will be an estimate of $\bra{\Psi_b} T_{\mathbf{v}} \ket{\Psi_b}$ accurate to $\pm O(1/\sqrt{S})$, provided $n,k,b$ are in the regime where the Markov chain converges. This is the same regime for which our quantum state preparation works and is characterized in \sect{sec:largeb}.

\section{Bounded Distance Decoding with Hadamard Tests}
\label{sec:bddimp}

To distinguish an overlap of $\epsilon$ from an overlap of zero via Hadamard tests requires order $1/\epsilon^2$ repetitions, as dictated by standard sampling statistics. Thus, $1/|\bra{\Psi_b} T_{\boldsymbol \delta} \ket{\Psi_b}|^2$ provides an estimate of runtime for this task as a function of $\delta = |\boldsymbol \delta|$. Neglecting sampling error, $|\bra{\Psi_b} T_{\boldsymbol \delta} \ket{\Psi_b}|$ is given by \eq{eq:ideal_overlap}. Given an empirical distribution over Hamming weights of dual codewords observed in our Monte Carlo simulations, we can also calculate $|\bra{\Psi_b} T_{\boldsymbol \delta} \ket{\Psi_b}|$ making only the assumption that dual codewords of given Hamming weight remain equiprobable. We omit this calculation, as it is much more involved than the calculation of the ideal overlap given in \sect{sec:overlap}.

\begin{figure}[hb!]
\begin{center}
\includegraphics[width=0.85\textwidth]{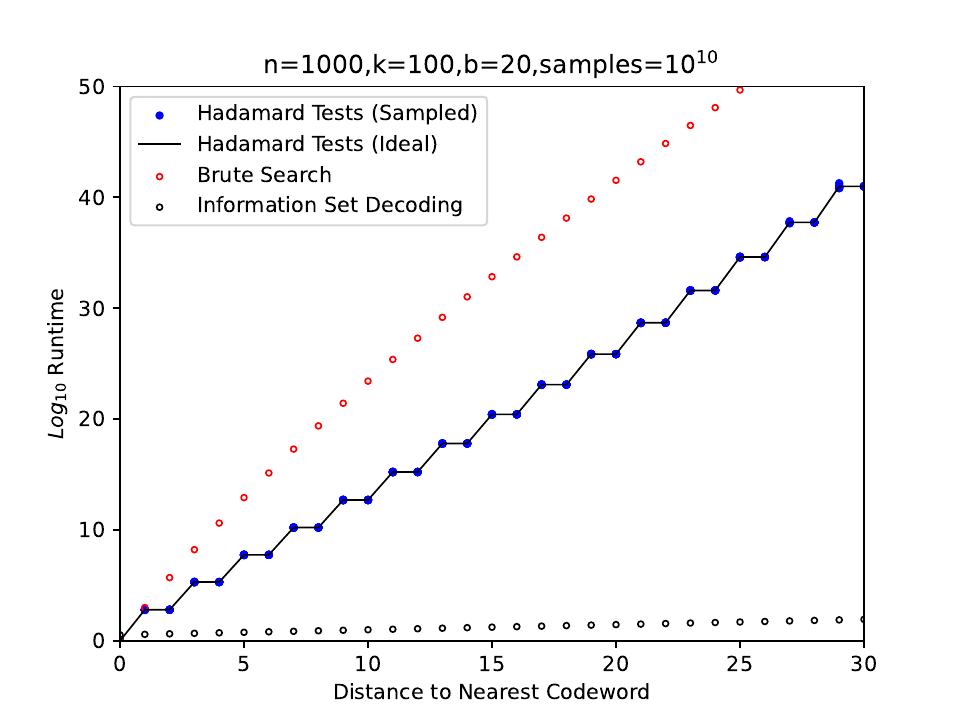}
\caption{\label{fig:runtimesgood}  The log of the runtimes for three approaches to Bounded Distance Decoding as a function of the distance to the nearest codeword. The red circles are from brute force search which the quantum algorithm beats but Information Set Decoding does the best. To illustrate the effect of sampling errors, the blue dots show the runtimes corresponding to the overlaps implied by the Hamming weight distributions on the dual code observed in ten Monte Carlo trials. This is for the well converged examples with $n=1000,k=100,b=20$. Consequently, the blue dots all lie on top of one another and match the ideal.}
\end{center}
\end{figure}

\begin{figure}[hb!]
\begin{center}
\includegraphics[width=0.85\textwidth]{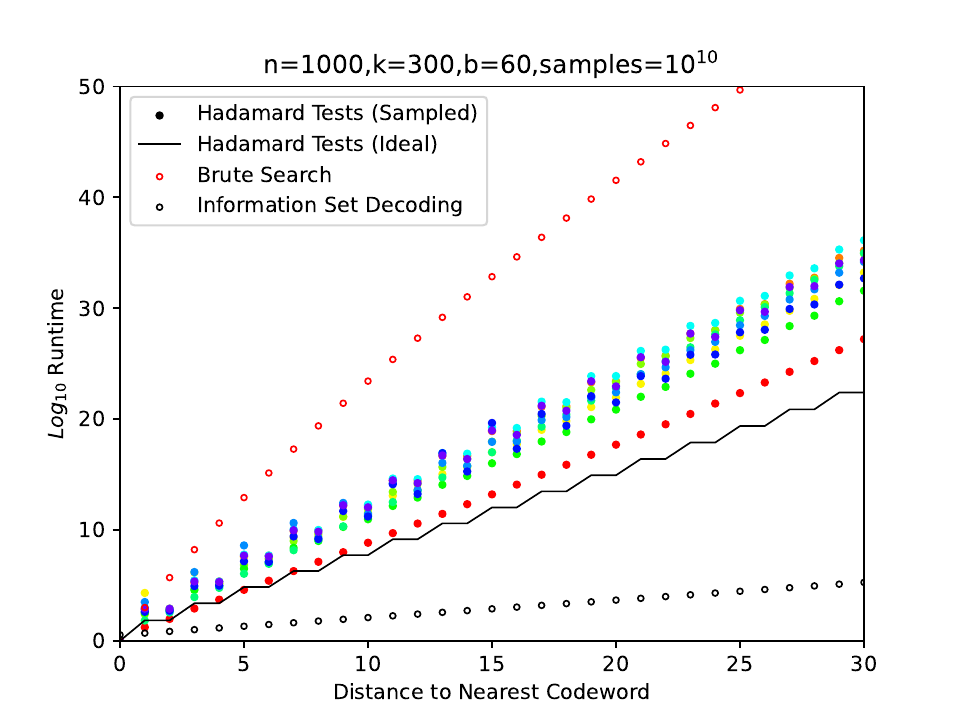}
\caption{\label{fig:runtimesbad}  The log of the runtimes for three approaches to Bounded Distance Decoding. Here $n=1000,k=300,b=60$ and the Monte Carlo converges incompletely. To illustrate the effect of sampling errors, we use ten colors to plot the runtimes corresponding to the overlaps implied by the Hamming weight distributions on the dual code observed in ten Monte Carlo trials. Even though the Monte Carlo has not perfectly converged the quantum algorithm beats brute force search in all ten cases.  Still Information Set Decoding beats the quantum algorithm.}
\end{center}
\end{figure}

But we show the runtimes in the presence of empirically observed sampling errors in figures \ref{fig:runtimesgood} and \ref{fig:runtimesbad} alongside the ideal runtimes achievable without sampling errors.

In figures \ref{fig:runtimesgood} and \ref{fig:runtimesbad} we compare these runtimes against classical brute force search, which has runtime $\binom{n}{\delta}$. We also compare against Information Set Decoding, a more sophisticated classical randomized algorithm described in \sect{sec:ISD}. We find that the quantum approach outperforms brute force search but falls short of Information Set Decoding. All three of these algorithms could be quadratically sped up using Grover search, leaving their relative performance unaffected. 

\clearpage

\section{Information Set Decoding}
\label{sec:ISD}

In this section we discuss a classical algorithm for BDD, specifically the most basic version of Information Set Decoding. This provides a useful point of reference because it is a simple classical algorithm that provides nearly state of the art performance. More recently, other more complicated classical algorithms have been devised that achieve slightly improved (but still exponential) scaling relative to ISD, \emph{e.g.} \cite{BLP11}. Some noteworthy features of ISD are that it uses only polynomial memory and it can be quadratically accelerated through Grover search \cite{B10}.

In a bounded distance decoding problem, we are given a generator matrix $B \in \mathbb{Z}_2^{k \times n}$ for a code $C$ and a vector $\mathbf{v}$ which is near an unknown codeword $\mathbf{c}$, that is, $\mathbf{v} = \mathbf{c} \oplus \boldsymbol \delta \in \mathbb{Z}_2^n$. Here $\boldsymbol \delta$ is the error vector satisfying $|\boldsymbol \delta| = \delta$ with $\delta$ not too large. So $\mathbf{u} B = \mathbf{c}$ for some $\mathbf{u} \in \mathbb{Z}_2^k$. The goal is to find $\mathbf{c}$, or equivalently $\mathbf{u}$.

The ISD algorithm works as follows.

\begin{enumerate}
    \item Select $k$ elements of $\{1,2,\ldots,n\}$ uniformly at random. Let $\mathbf{v}^{(k)}$ be the length-$k$ string obtained by keeping only these bits from $\mathbf{v}$ and let $B^{(k)}$ be the $k \times k$ submatrix of $B$ obtained by keeping only these columns of $B$.
    \item Check whether $B^{(k)}$ is an invertible matrix over the field $\mathbb{Z}_2$. If not, go back to step 1 and try a different subset.
    \item Solve the linear system $\tilde{\mathbf{u}} B^{(k)} = \mathbf{v}^{(k)}$ for $\tilde{\mathbf{u}} \in \mathbb{Z}_2^k$ using Gaussian elimination.
    \item Compute the Hamming distance $| \tilde{\mathbf{u}} B \oplus \mathbf{v} |$. If this is equal to $\delta$ then $\tilde{\mathbf{u}} B$ is a codeword of distance $\delta$ from $\mathbf{v}$ and we are done. Otherwise, go back to step 1 and try again with a different subset.
\end{enumerate}

Next, let us analyze the probability of success for each trial. The probability that a uniformly random $k \times k$ matrix over the field $\mathbb{Z}_2$ is invertible, in the limit of large $k$, is approximately 0.29 \cite{W87}. For generic dense $B$ this is also an accurate prediction of the approximate probability that the random submatrix $B^{(k)}$ will be invertible in step 2.

Since $\mathbf{v} = \mathbf{c} \oplus \boldsymbol \delta$, the linear system we solve in step 3 can be rewritten as
\begin{equation}
    \tilde{\mathbf{u}} B^{(k)} = \mathbf{c}^{(k)} \oplus \boldsymbol \delta^{(k)}.
\end{equation}
If we are lucky, the subset of bits we chose contains no errors. That is $\boldsymbol \delta^{(k)} = \mathbf{0}$. In this case,
\begin{equation}
    \label{eq:goodsubset}
    \tilde{\mathbf{u}} B^{(k)} = \mathbf{c}^{(k)}.
\end{equation}
Next, recall that since $\mathbf{c}$ is a codeword there is a $\mathbf{u}$ such that $\mathbf{u} B = \mathbf{c}$. Given $\mathbf{c}$ one could solve this system of linear equations to obtain $\mathbf{u}$. In fact, this system of equations is overdetermined, having $n$ constraints but only $k$ unknowns. Throwing away all but $k$ of these constraints will still provide a fully determined system of equations, provided that the remaining submatrix of $B$ is invertible. Since 
 $\boldsymbol \delta^{(k)}=\mathbf{0}$, the system we solve in step three is \eq{eq:goodsubset}, which is precisely this same system of equations, which we can solve to obtain the unique solution $\tilde{\mathbf{u}} = \mathbf{u}$. Now $\tilde{\mathbf{u}} B$ is $\mathbf{c}$, which we have checked in step four has Hamming distance $\delta$ from $\mathbf{v}$ and so the algorithm terminates with success. How likely is this to happen?

The string $\boldsymbol \delta$ has $n$ bits, of which $\delta$ have errors and each trial randomly selects $k$ of these $n$ bits. The probability that the subset contains no errors is exactly
\begin{equation}
    \frac{\binom{n-\delta}{k}}{\binom{n}{k}}.
\end{equation}
Hence, the probability of success in one trial is
\begin{equation}
    p_{\mathrm{ISD}} \simeq 0.29 \frac{\binom{n-\delta}{k}}{\binom{n}{k}}.
\end{equation}
If $\delta$ is small compared to $n$ and $k$ then one can make the approximation
\begin{equation}
    p_{\mathrm{ISD}} \simeq 0.29(1-k/n)^\delta.
\end{equation} 
The expected runtime of ISD is $1/p_{\mathrm{ISD}}$, which can be reduced to $1/\sqrt{p_{\mathrm{ISD}}}$ if one replaces random sampling with Grover search \cite{B10}.

\section{Discussion}

In this work, we show how to solve the Quantum Distributional State Construction problem for $\ket{\Psi_b}$, the superposition over bit strings within Hamming distance $b$ of a linear code of length $n$ and dimension $k$. We map out the region of $n,k,b$ where this procedure efficiently yields a high fidelity approximation to $\ket{\Psi_b}$. We find that this region extends beyond the uncomputation barrier that thwarts a direct approach to constructing $\ket{\Psi_b}$.

We restricted our investigations in several respects. We considered only balls whose radius is less than half the distance of the code, so that the balls do not overlap. We considered only random binary linear codes. And the only application we considered is the bounded distance decoding problem where we did not beat a known classical algorithm  and our procedure can be dequantized. Given that $\ket{\Psi_b}$ can be constructed efficiently out to surprisingly large $b$, it may be interesting to consider potential algorithmic, cryptographic, or complexity-theoretic applications beyond what we considered here. Or, the state preparation method used here could perhaps be applied to overcome uncomputation barriers in other Quantum Distributional State Construction problems.\\
\\
\noindent
\textbf{Acknowledgments: } We thank Madhu Sudan for discussions a long time ago. We thank Noah Shutty and Bill Huggins for more recent discussions. And we particularly thank Oded Regev for a key observation.

\appendix

\section{Phase Kickback}
\label{app:kickback}

Let $f:\{0,1\}^n \to \{0,1\}$ be a function for which we know an efficient classical circuit. A unitary $U_f$ can always be implemented efficiently such that \cite{NC00}:
\[
    U_f \ket{\mathbf{x}} \ket{z} = \ket{\mathbf{x}} \ket{z \oplus f(\mathbf{x})}
\]
where the second register is a single qubit.
By initializing the second register to $\ket{0}$ one obtains $f(\mathbf{x})$ written into the computational basis in the second register.  One can instead compute $f$ into the phase by initializing the second register with $\frac{1}{\sqrt{2}} \left( \ket{0} - \ket{1} \right)$ so
\[
    U_f \ket{\mathbf{x}} \frac{1}{\sqrt{2}} \left( \ket{0} - \ket{1} \right) = (-1)^{f(\mathbf{x})} \ket{\mathbf{x}} \frac{1}{\sqrt{2}} \left( \ket{0} - \ket{1} \right).
\]
The second register remains unentangled with the first and can be discarded. The net effect is $\ket{\mathbf{x}} \to (-1)^{f(\mathbf{x})} \ket{\mathbf{x}}$. By linearity, an arbitrary superposition undergoes the transformation
\[
    \sum_{\mathbf{x}} a(\mathbf{x}) \ket{\mathbf{x}} \to \sum_{\mathbf{x}} (-1)^{f(\mathbf{x})} a(\mathbf{x}) \ket{\mathbf{x}}.
\]
This is known as phase kickback and  it is  one of the  steps in the state preparation algorithm.

\section{Systematic Form for Linear Codes}
\label{app:systematic}

The $k \times n$ generator matrix $B$ for a code $C$ is not unique. The basis vectors $\mathbf{b}_1,\ldots,\mathbf{b}_k$ that form the rows of $B$ can be replaced by any invertible $\mathbb{Z}_2$-linear combination of $\mathbf{b}_1,\ldots,\mathbf{b}_k$ without changing the code. One consequence of this is that $B$ can be put into \emph{systematic form}:
\[
    B = \left[ \id_{k \times k} | R \right],
\]
where $\id_{k \times k}$ is an identity matrix and $R$ is a general $k \times (n-k)$ matrix. Thinking of $\mathbb{Z}_2$ as a finite field, systematic form is recognizable as reduced row echelon form, which can be obtained in $O(k^2 n)$ classical operations using Gaussian elimination. Let
\[
    B_{\perp} = \left[ R^T | \id_{n-k\times n-k} \right].
\]
By construction, $B B_\perp^T = R \oplus R = 0$. Hence, $B_\perp$ is a generator matrix for the dual code $C^\perp$. From this construction, we see that given a generator matrix for $C$, a generator matrix for $C^\perp$ can be obtained in polynomial time classically.   Also if we want to generate a random code and its dual at the same time we can just generate a random $k \times (n-k)$ matrix $R$ and then get $B$ and $B_\perp$ immediately.

\section{Krawtchouk Identity}
\label{app:ident}

As stated in \eq{eq:ident}, Krawtchouk polynomials obey the identity
\[
\sum_{j=0}^b K_j^n(x) = K_b^{n-1}(x-1).
\]

To show this, we note that binomial coefficients have the property $\binom{x}{r} = \binom{x-1}{r} + \binom{x-1}{r-1}$. Thus
\begin{eqnarray*}
\sum_{j=0}^b K_j^n(x) & = & \sum_{j=0}^b \sum_{r=0}^j \binom{x}{r} \binom{n-x}{j-r} (-1)^r \\
& = & \sum_{j=0}^b \sum_{r=0}^j \left[ \binom{x-1}{r}+\binom{x-1}{r-1} \right] \binom{n-x}{j-r} (-1)^r \\
& = & \sum_{j=0}^b \left( \left[ \sum_{r=0}^j \binom{x-1}{r} \binom{n-x}{j-r} (-1)^r \right] + \left[ \sum_{r=0}^{j-1} \binom{x-1}{r} \binom{n-x}{j-r-1} (-1)^{r+1} \right] \right) \\
& = & \sum_{j=0}^b \left( K_j^{n-1} (x-1) - K_{j-1}^{n-1}(x-1) \right) \\
& = & K_b^{n-1}(x-1),
\end{eqnarray*}
where the last line follows from $K_{-1}^{n-1}(x-1) = 0$. $\square$

\bibliography{construction}

\end{document}